\documentstyle[prb,aps,epsf]{revtex}
\begin{document}

\widetext

\title{Charge ordering and long-range interactions in layered transition
metal oxides: a quasiclassical continuum study }
\author{Branko P.\ Stojkovi\'c,$^{1}$ Z.\ G.\ Yu,$^{2}$ A.\ L.\
Chernyshev,$^{3}$\cite{perm} 
A.\ R.\ Bishop,$^{1}$ A.\ H.\ Castro Neto$^{3}$
and Niels Gr{\o}nbech-Jensen$^{4}$}

\address{$^{1}$Theoretical Division and Center for Nonlinear Studies, Los
  Alamos National Laboratory, Los Alamos, New Mexico 87545}
\address{$^2$Department of Chemistry, Iowa State University, Ames, IA 50011}
\address{$^{3}$Department of Physics, University of California, Riverside, CA
  92521}
\address{$^4$Department of Applied Science, University of California,
  Davis, California 95616 \\
  and\\
  NERSC, Lawrence Berkeley Laboratory,
  Berkeley, California 94720}
\date{\today}
\maketitle
\begin{abstract}
  The competition between long-range and short-range interactions
  among holes
  moving in an antiferromagnet (AF), 
  is studied within a model derived from the
  spin density wave picture
  of layered transition metal oxides.  A novel numerical approach is
  developed which allows one to solve the problem at finite hole
  densities in very large systems (of order hundreds of lattice
  spacings), albeit in a quasiclassical limit, 
  and to correctly incorporate the long-range part of the
  Coulomb interaction. The focus is on the problem of
  charge ordering and the charge phase diagram: at low temperatures
  four different phases are found, depending on the strength of the
  magnetic (dipolar) interaction
{generated by the spin-wave exchange}, 
  and the density of holes. The four phases are the
  Wigner crystal, diagonal stripes, a grid phase (horizontal-vertical
  stripe loops) and a glassy-clumped phase. In the presence of both
  in-plane and out-of-plane charged impurities the stripe ordering is
  suppressed, although finite stripe segments persist. At finite
  temperatures multiscale (intermittency) dynamics is found,
  reminiscent of that in glasses. The dynamics of
  stripe melting and its implications for experiments is discussed.
\end{abstract}
\phantom{.}

\section{introduction}
\label{sec:intro}

Charge ordering in layered transition metal oxides has recently attracted a
significant research interest, due to its possible relation to the mechanism
of high temperature superconductivity in doped cuprates\cite{zachar} and
bismuthates.\cite{bismuthates}  In particular,
stripe-like ordering, which involves holes ordered into linear arrays,
separated by an antiferromagnetically (AF) ordered electronic
background, have been discussed as a candidate for the explanation of
pseudogap effects in underdoped cuprate compounds.\cite{zachar}  
In addition, the formation
of domain walls has been discussed in terms of the proximity to phase
separation.\cite{ps} Quite generally, phase separation on mesoscopic
and even macroscopic scales is potentially relevant for any strongly
correlated organic and inorganic 
electronic system, including {systems with} 
spin-density-wave (SDW),\cite{sdw_ordering}
charge-density-wave (CDW),\cite{cdw_ordering} and Jahn-Teller
broken-symmetry\cite{jahn_teller} ground states.  

On the experimental side, mesoscopic (nanoscale) 
phase separation has been observed in
many compounds. 
In the case of La$_{2-x}$Sr$_x$NiO$_{4+y}$ stripes have been observed
both using nuclear magnetic resonance (NMR) methods and 
more directly, using high-resolution
electron diffraction \cite{nickel}. 
In addition, 
stripes have also been identified in La$_{1-x}$Ca$_x$MnO$_3$
for specific commensurate values of doping \cite{stripemagneto}.
In cuprates static stripe order has been observed in 
La$_{1.6-x}$Sr$_x$Nd$_{0.4}$CuO$_4$ in both elastic and inelastic
neutron scattering experiments\cite{tranquada}, and $x$-ray diffraction
experiments \cite{xray}. 
{There are also evidences that stripes exist in some form in
high-$T_c$ compounds.}
In the oxygen doped
La$_{2}$CuO$_{4+\delta}$\cite{hammel} stripes have been observed
using the nuclear magnetic resonance (NMR) techniques. 
Magnetic susceptibility measurements\cite{suscep}, nuclear quadrupole
resonance \cite{nqr} (NQR)
and muon spin resonance \cite{msr}
all indicate formation of domains in La$_{2-x}$Sr$_x$CuO$_{4}$
and recent inelastic neutron scattering (INS) experiments
in La$_{2-x}$Sr$_x$CuO$_{4}$ and YBa$_2$Cu$_3$O$_{7-\delta}$
superconductors yield results consistent
with stripe formation \cite{LaSr,agree,aeppli},
although the width of the INS lines in, e.g., 
YBCO materials is large, which may suggest dynamic charge ordering.

On the theoretical side, stripes have been proposed by several
research groups.  Since in strongly correlated systems, such as
cuprate superconductors, electrons exhibit a strong on-site repulsion,
numerous studes have been devoted to the Hubbard and $t$-$J$ models.
It has been shown that a mean-field treatment of the Hubbard model
yields a stripe phase as a locally stable solution \cite{hartree-fock}.
Many other studies view the stripes
as an outcome of the competition between kinetic energy of holes and
exchange energy of spins alone and {frequently} neglect % 
the role of the
long-range part of the Coulomb interaction.\cite{subsequent,C_CN_B} and only 
recently, an attempt to incorporate the long-range forces into the
mean-field approach to the Hubbard model has been made.\cite{Coulomb}
Another point of view emphasizes the intrinsic instability of a
strongly correlated electronic system towards a phase separation as a
necessary starting point.\cite{emery1,emery2} 
Then it is assumed that such
an instability is prevented by the long-range Coulomb forces. Therefore, the
competition between this instability, whose existence in the physical
range  of parameters of the realistic models is yet to be proven, and
Coulomb repulsion gives rise to a stripe phase. Thus, these two
approaches agree on the importance of the correlations but disagree on
the role of long-range forces. More recently, it has
been shown that phase separation is indeed a very common phenomenon
close to quantum critical points.\cite{paco}

One would expect that the existence of stripes in the widely studied
``minimal'' $t-J$ or 
Hubbard models can be either proven or 
disproven by some unbiased numerical technique.
Unfortunately, numerically the stability of the stripe phase has been
established 
less clearly.  Numerical studies of the $t$-$J$ model are presenting
conflicting conclusions as to the existance of stripe phases in the
ground state of this ``basic'' strongly correlated model which might
be the result of the strong finite-size effects \cite{white,HM}.
For example, even a Monte Carlo simulation of the doped Ising model,
without the long-range forces, yields holes ordered into
loops, rather than into geometric arrays.\cite{Zaanen_private} 
In fact, with an exception of the recent Density Matrix Renormalization
Group (DMRG) simulations of White and Scalapino\cite{white}, which
have found stripe formation in relatively large $t-J$ clusters, no
microscopic calculation to date has shown that the stripes are a
stable entity.  Most importantly, no stripe formation in a system with
long-range interaction has been {studied}
in a direct simulation. 
In addition, the sizes of the clusters available with modern day
computers for solving quantum models of spins and holes are still too small 
to study role of the the long-range Coulomb interaction and 
be free from significant finite-size effects. 

In this situation we propose a different strategy:
one can study a quasiclassical limit of the quantum problem of holes
in an AF spin environment analytically
and incorporate all essential correlations in
an effective hole-hole interaction. {In this case
the AF background is effectively integrated out, and
the focus is on the charge subsystem.}
Then the motion of 
``classical'' holes at finite density, 
interacting via an \emph{effective} magnetic interaction and in the
presence of long-range Coulomb forces,
can be studied numerically in much larger systems. In other words, in
this paper we combine analytical and numerical approaches to study the
charge ordering in transition metal oxides.

Our numerical approach is based on the spin density wave (SDW)
picture of Schrieffer, Wen and Zhang,\cite{bob} which is 
{closely related}\cite{david} 
to the 
{semiclassical approach to the $t$-$J$ model by}
Shraiman and Siggia\cite{ss} in which 
the interaction between doped holes stems from the spiral distortion
of the local N\'{e}el vector near a hole. As
shown below, in the quasiclassical limit, % 
the problem can be solved using classical Monte Carlo (MC) or
molecular dynamics (MD) methods. In a systematic numerical study
we explore
the interplay between long-range Coulomb interaction and
short (or intermediate)-range AF interactions of dipolar nature,
which we take to have both
isotropic and anisotropic components (depending on the lattice structure).

Our main results can be summarized as follows: in the absence of
disorder we find four phases depending on the density of holes and the
characteristic AF energy scales: ({\em i}) a Wigner crystal, ({\em
  ii}) diagonal (glassy)
stripes, ({\em iii}) a geometric phase, 
characterized by horizontal-vertical stripes
or checkerboard (grid), and ({\em iv}) 
a ``clumped'' phase (phase separation).
In our study the stripe-like phases emerge as a kind
of melting of the Wigner crystal phase, hence the long-range Coulomb
interaction is a necessary ingredient for their occurrence.
In the geometric phase the stripes, resulting from the competition of
the short-range and long-range interactions, are characterized by a
particular AF dipolar alignment. The patterns are very stable, showing 
large ``string tension,'' while the motion of holes within a stripe is
much softer. If one takes into account the kinetic energy
of the holes along the stripes one is lead to the concept of
a quantum liquid crystal as proposed recently by Emery, Fradkin
and Kivelson. \cite{efk}
On the other hand, the ground state of the geometric phase
is not well defined in that there are many geometric phases with
very low energies, comparable to that of the ground state, implying a
{\em rugged energy landscape}. We find
that, on lowering the temperature, the geometric hole ordering is
characterized by occurence of secondary defects in the structure. At higher
temperatures we find that the {\em dynamic} hole ordering is
characterized by {\em temporally intermittent} pattern formation
(i.e., spatio-temporal intermittency).
Finally, we find that a sufficient concentration of randomly placed
impurities destroys the geometric hole pattern, although, regardless of
the impurity type, stripe {\em segments} are preserved.

The paper is organized as follows: in the next section we present a
review of the theoretical model and the computational methods we use.
In section \ref{sec:results} we present our numerical results and in
particular we present the phase diagram showing how the obtained
phases emerge as a function of doping and interaction strengths.
Finally, in section \ref{sec:conclusions} we summarize our
conclusions and experimental implications.

\section{model}
\label{sec:model}

We begin with the spin density wave (SDW)
 picture of layered transition
metal oxides. This picture has been very successful in describing the
stoichiometric insulating AF phase of these systems at low
 temperatures.\cite{slater,bob} 
In this picture the electrons move with hopping
energy $t$ in the
self-consistent staggered field of its spin, as described by, e.g.,
 the Hubbard model (the calculation is presented in detail in Appendix
\ref{hub}):
\begin{equation}
H= -t\sum_{\langle i,j \rangle} (c_i^\dagger c_j + h.c.) +U\sum_i
n_{i,\uparrow} n_{i,\downarrow}.
\label{eq:hubbard}
\end{equation}
Because the
translational symmetry of the system
is broken, the electronic band is split
into upper and lower Hubbard bands.\cite{mott}
On performing  a Bogoliubov transformation,  one defines 
the valence ($h_{k,\alpha}$) and conduction band ($p_{k,\alpha}$)
operators,  respectively:
\begin{eqnarray}
p_{k,\alpha} &=& u_k c_{k,\alpha} + \alpha v_k c_{k+Q,\alpha}\\
h_{k,\alpha} &=& u_k c_{k,\alpha} - \alpha v_k c_{k+Q,\alpha},
\label{eq:vc}
\end{eqnarray}
where $u_k$, $v_k$ are the Bogolioubov weights and $\alpha$ is the
sublattice index. 
The upper and lower Hubbard bands are
separated by the Mott-Hubbard gap, $\Delta=US/2$, where $S$ is the
expectation value of the staggered field $S_z$,
\begin{equation}
\langle S_z({\bf q})\rangle = -2 \sum_{k,\alpha} u_{k+q-Q} v_k \langle
 h_{k+q-Q,\alpha} h^\dagger_{k,\alpha}
\rangle ,
\label{eq:staggered}
\end{equation}
calculated at momentum transfer
${\bf q}={\bf Q}$. At half
filling the lower band is filled and the upper band is empty.
This picture is consistent with the angle resolved photoemission data
in the layered AF insulator Sr$_2$CuO$_2$Cl$_2$.\cite{photo}
On doping the system with holes with planar density $\sigma_s$,
at low temperatures, $T \ll \Delta/k_B$, the low frequency physics 
reflects  purely the lower
Hubbard band (LHB). It has been shown\cite{andrey} that, regardless of
the band structure, the LHB has
 a maximum at four wavevectors ${\bf k}_i = (\pm 1,\pm 1)\pi/(2 a)$,
where $a$ is the lattice spacing, and therefore the long wavelength
theory of the problem can be studied by assuming the momentum of
the holes to be close to these points.

Then the two hole interaction Hamiltonian
can be
separated into the longitudinal and transverse parts ($H_z$ and $H_{xy}$,
respectively), whose Fourier transform, for quasiparticle momenta near $k_i$,
 is equal to:\cite{david}
\begin{eqnarray}
\label{eq:interaction}
H({\bf r}) &&= \left[ A_z\sigma^z_1 \sigma^z_2 - 
 A_{xy}\left(\sigma^+_1\sigma^-_2 + \sigma^-_1\sigma^+_2
\right)\right] \delta({\bf r})- \nonumber \\
B_{xy}&&\left[\frac{{\bf d}_1\cdot {\bf d}_2}{r^2} - 
2\frac{({\bf d}_1 \cdot {\bf r})\,\, 
        ({\bf d}_2 \cdot {\bf r})}{r^4}\right]
\left(\sigma^+_1\sigma^-_2 + \sigma^-_1\sigma^+_2\right),
\end{eqnarray}
{where $r=|{\bf r}_1-{\bf r}_2|$ is a relative hole-hole distance,
${\bf r}_i$ is a coordinate
of a hole in units of $a$,
$\sigma^{z(\pm)}_i=c^{\dag}_{\alpha}\sigma^{z(\pm)}_{\alpha\beta}c_{\beta}$
is a spin-density operator, with $\sigma^{z(\pm)}$ Pauly matrices,
${\bf d}_i$ is a unity vector in the direction of the dipole moment of
the hole.} In the SDW formalism the interaction strengths
$A_z$, $A_{xy}$ and $B_{xy}\sim A_{xy}/2\pi$ are all of order Hubbard
$U$. The interaction (\ref{eq:interaction}) is clearly rotationally
invariant and valid for $r\geq a$, while
for $r\rightarrow 0$ it yields an unphysical divergence of the
(attractive) dipolar interaction.
We observe that this form of the Hamiltonian is not particular to
the SDW theory of the Hubbard model, but stresses the fact that
a mobile carrier in an antiferromagnet produces a dipolar distortion
of the magnetic background. We demonstrate this explicitly in Appendix
\ref{tj} where we show that the t-J model has exactly the same type of
interaction terms.
In other words, at finite density 
 the holes interact via two different mechanisms:
a uniform short-range attractive force due to AF bond--breaking 
and a long-range magnetic dipolar interaction 
(contained by 
Eq.\ (\ref{eq:interaction}), see Ref.\ \onlinecite{bob}). 
The latter term is due to the long range spiral
distortion of the AF background, which is a consequence of
 quasiparticles interacting with soft (Goldstone) modes of the
spin  system.\cite{david,ss}
The magnetic dipole
moment associated with each hole is due to the coherent
hopping of holes between different sublattices and scales with the
AF magnetic energy. {This implies that the quantum effects
  associated with hole kinetic energy can be neglected, which is
  correct in the limit $t\ll J$, believed to be valid in nickelates.
  This is also why the} hole-hole interaction obtained in
the weak coupling SDW picture is
equivalent\cite{david} to that {in the effective Hamiltonian} found 
{by} Shraiman and
Siggia\cite{ss}, based on the $t-J$ model, where the \emph{dipolar}
interaction is obtained using \emph{semiclassical} analysis of the spin
part of the model, as well as symmetry considerations. 
It is also possible to prove,
using Ward identities, that the remaining spin part of the
problem is equivalent to the two dimensional (2D) 
non-linear $\sigma$ model in
the long wavelength limit \cite{fradkin}. 
It has been argued\cite{ss}
that at physical values $t/J\gg 1$ all coupling strengths ($A_z$,
$A_{xy}$ and $B_{xy}$), and
therefore  the hole-hole interaction, will be renormalized to the
value of super-exchange constant $J$.

In pure two dimensions at finite $T$ the system is magnetically
disordered, characterized by a finite magnetic 
correlation length, $\xi$ (see Ref.\ \onlinecite{chak}), and
the range of the dipolar interaction between the holes,
mediated by the AF background, is also of order $\xi$.
In fact, even at $T=0$ and  \emph{finite} hole concentration the
correlation lenght is restricted and
the dipolar interaction is effectively short ranged.

It has been noted \cite{SK} that the dipolar twist of the magnetic
background would imply local time-reversal symmetry breaking which puts
some restrictions on the applicability of the picture to the real
systems where such symmetry breaking has not been experimentally
observed. Indeed, the time-reversal symmetry is already broken in
the N\'eel state, as well as in any magnetically ordered state. 
However, 
in two-dimensional spin systems at finite temperature all symmetries
are restored and we expect this to be true also for the hole-doped case
studied here.

Besides the AF interactions the holes also experience the long-range
Coulomb interaction. This is clear if we consider that $r_s=r_0/a_0$
(where $r_0$ is the mean inter-particle distance and $a_0$ is the
Bohr radius) is very large in the underdoped systems ($r_s \approx 8$).
In other words, in systems with a small density of holes the screening,
which is 
due to the density fluctuations, is very weak and one must take the
Coulomb interaction into account.

Finally, each hole carries a spin degree of freedom as well. However
inspection of Eq.\ (\ref{eq:interaction}) reveals that the
overall spin energy is minimized in
the spin anti-symmetric channel. Hence we neglect the spin-symmetric
channel and thus
 in our approach, we only consider  the charge channel with an
effective (magnetic in origin)
 interaction between two holes, $1$ and $2$, of the form
(see Eq.\ (\ref{eq:interaction}))
\begin{eqnarray}
V({\bf r})= \frac{q^2}{\epsilon r}
-A \, e^{-r/a}-B \cos (2\theta-\phi_1-\phi_2){e^{-r/\xi}} % 
.\label{total}
\end{eqnarray}
{where we have asssumed that {\bf r} can be relaxed
from a crystal lattice position to an
arbitrary (continuous) value. We return to this point later.}

{In Eq.\ (\ref{total})}
$q$ is the hole charge, $\epsilon$ is the dielectric constant
(which we assume to be of order 1), $\theta$ is the angle made between
${\bf r}$ and a fixed axis and $\phi_{1,2}$ are the
angles of the magnetic dipoles
relative to the same fixed axis. $A$ is the strength of the uniform
(short-range) interaction
and $B$ is the strength of the magnetic 
dipolar interaction, 
which we will assume to be independent 
adjustable parameters, which, in real materials, should be of order
$\sim 1$ eV.
Note that we have introduced $B\sim B_{xy}/l^2$, where $l$ is some
appropriate average length, $a<l<\xi$, in order to avoid the
unphysical divergence of the dipolar part of the interaction in Eq.\ 
(\ref{eq:interaction}), while keeping the necessary symmetry of the
interaction. Moreover,
in the SDW picture, at low doping, $\phi_1$ and $\phi_2$ are
restricted to the angles $(2n+1)\pi/4$, the four angles determined by
the vectors {\bf k}$_i$. {In addition, the dipole moment vectors
are also restricted to {\bf k}$_i$, which justifies our use of Eq.\
(\ref{total}) where we have assumed a fixed \emph{size} 
of the dipole moment for each hole.}
However, at larger doping levels these angles
can be relaxed to arbitrary values, provided the interaction is
short-ranged. Of course, we have verified by an explicit calculation
that restricting dipole angles to the discrete values 
does not qualitatively change our results,
presented in the next section.
It is interesting to note that the hole-hole interaction in  the
form almost identical to Eq. (\ref{total}) has been obtained by
Aharony {\it et al.}\cite{Aharony} for the {\it static} holes residing
on Cu-O bonds within the framework of a {\it classical} model of an AF
diluted by ferromagnetic bonds. In this case the value of the coupling
constant $B$ is also restricted by a few $J$. Indeed, starting from
the insulating
phase of cuprates, the holes are injected into the CuO$_2$ planes
at high temperatures during the sample preparation. The 
hole distribution in this case is annealed {(instead of quenched as
proposed in Ref.\ \onlinecite{Aharony})} since the holes have
enough phase space for interactions. As the temperature is lowered
the holes can adjust themselves to $V({\bf r})$ and form the
structures we discuss below.

Quite generally, one can think of an AF as an active media
generating long range dipolar forces in response on some local
distortion. Therefore, the interaction $V({\bf r})$, Eq. (\ref{total}) is of
more general significance than just a result of the SDW picture, and the
study of the system of particles interacting via  $V({\bf r})$ is of wider
interest.

In general, the many-body problem of holes in an AF
background is extremely complicated, involving many-particle
interaction terms. However, at low densities, where the average
distance between holes is comparable to the AF correlation length, it is
reasonable to assume that the interaction of any two holes is weakly
perturbed by other holes, and the total potential energy can be
expressed in terms of two-particle energies, provided the Af
correlation length is replaced by an \emph{effective} correlation
length, which, to avoid clutter, we also denote as $\xi$. We therefore study
the system of ``classical'' particles in a computational box of size
$L_x\times L_y$, interacting via a potential
\begin{equation}
H_I=\sum_{<i,j>} V({\bf r}_{ij}),
\label{eq:hamiltonian}
\end{equation}
where $V({\bf r})$ is given by (\ref{total}).
However, we
emphasize that our approach is not a self-consistent one in the sense
that the true interaction must include many particle terms (omitted
here), which stem
from the fact that the SDW state is altered due to the charge
ordering.
The self-consistent approach to charge ordering will be presented 
elsewhere. {In addition, superconducting fluctuations have
  been neglected.} 
Moreover, the kinetic energy of the holes may lead to a 
quantum melting of the phases discussed here.

\begin{figure}
\centerline{
{\epsfxsize=7.5cm \epsfbox{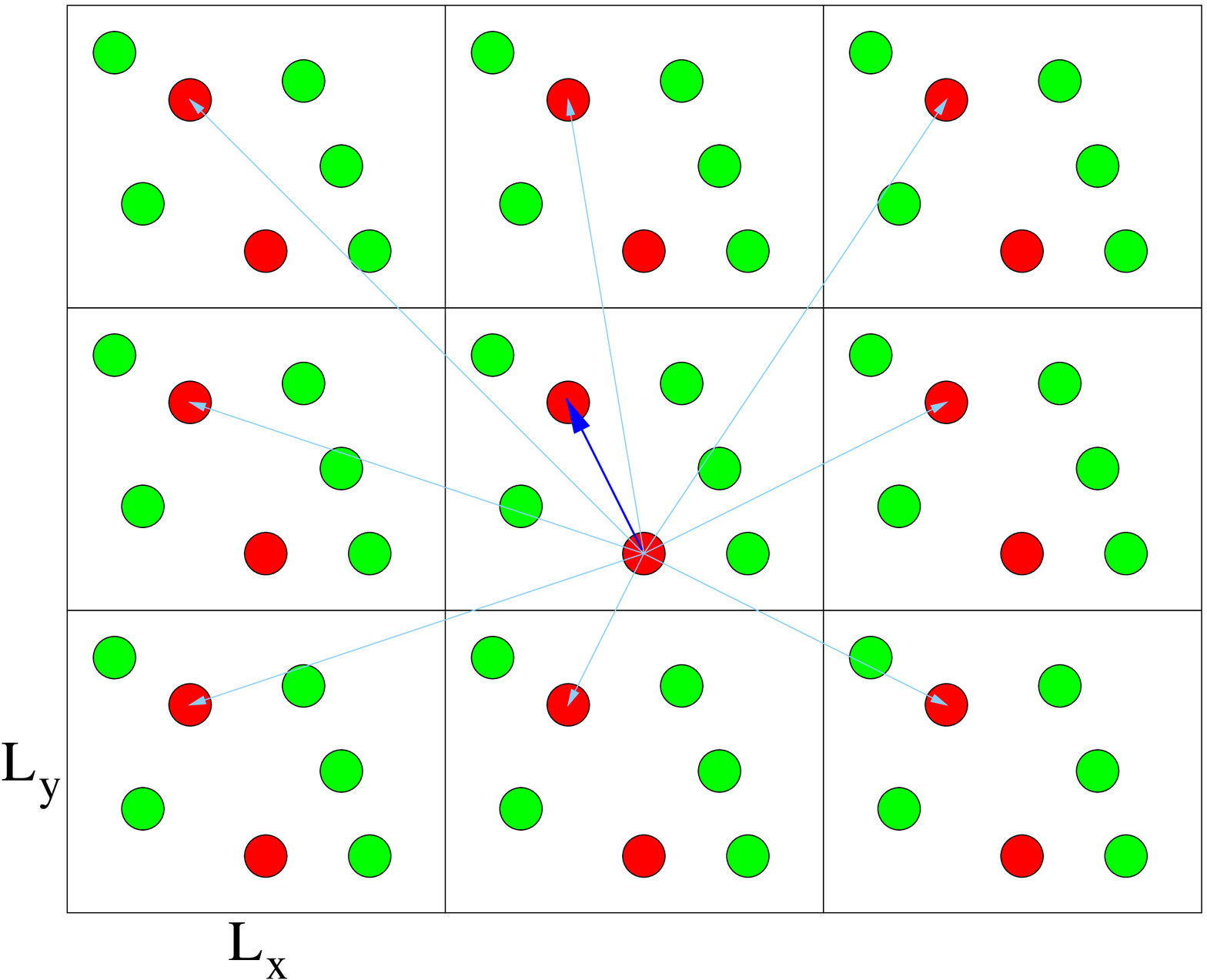}}
}
\centerline{
{\epsfxsize=7.50cm \epsfbox{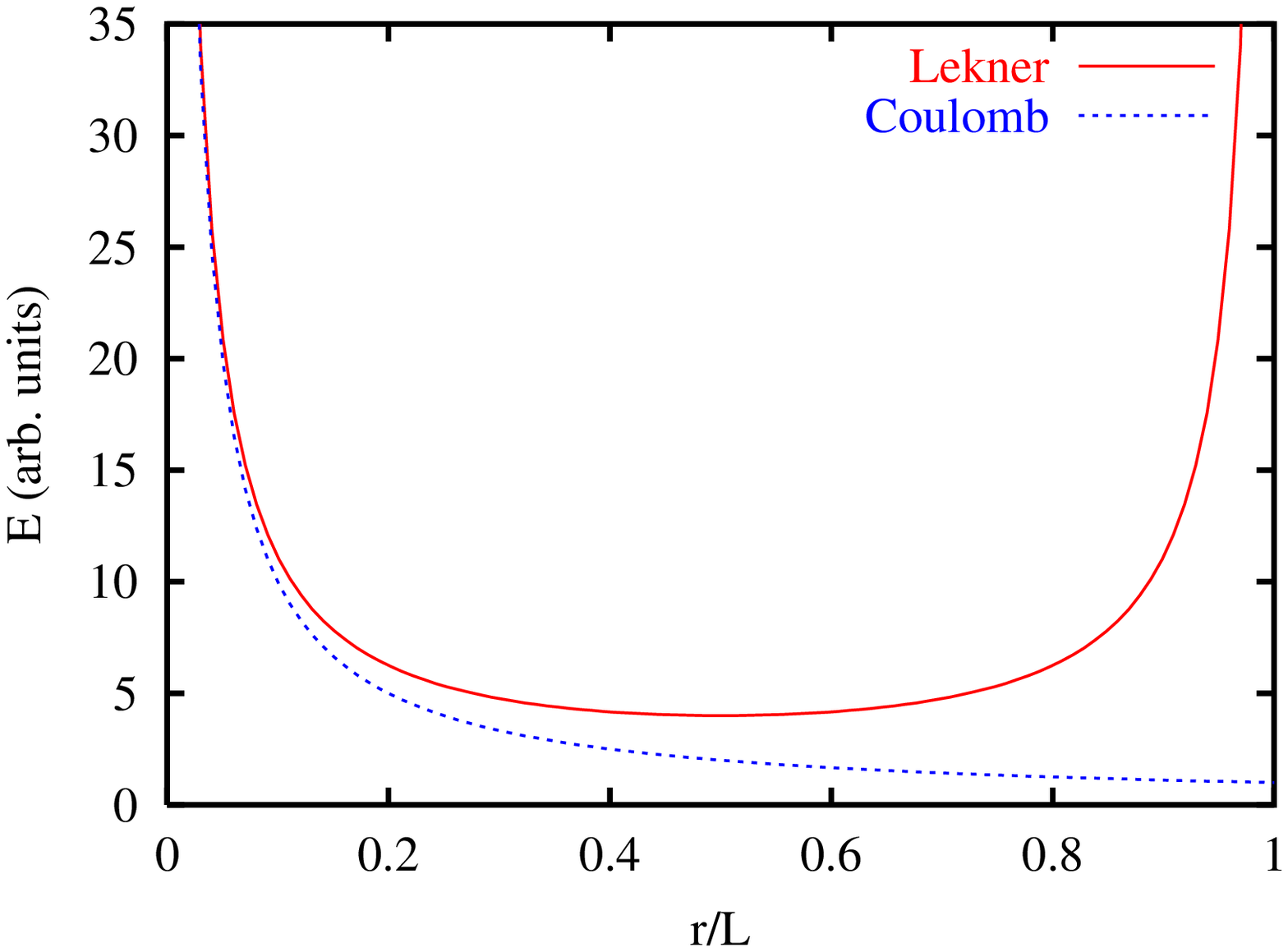}}
}
\vskip1cm
\caption{Top: Electrostatic interaction in a system with periodic
  boundary conditions: an effective interaction between any two charged
  particles in the computational box (central rectangle), such as the
  two marked by a thick vector line, involves the interaction with all 
  of the particle images, marked by thin wave vector lines. Bottom:
The Lekner potential (which accounts for the periodic boundary
conditions) in comparison with the bare Coulomb potential, within a
computational box of size $L$. As expected, at $r\rightarrow 0$ the
two potentials coincide (up to an additive 
constant) and hence their difference
can be readily approximated by a low order polynomial.}
\label{fig:lekner}
\end{figure}
The long-range character of the Coulomb interaction requires further
consideration:  in order to perform calculations at
finite density, as required by the dipolar interaction we consider, we
usually perform calculations with periodic boundary conditions.
The ability to handle long-range Coulomb
interactions in rectangular periodic media
has been enhanced recently in the area of molecular
physics.\cite{niels} 
Assuming a computational cell of arbitrary
geometry and cyclic boundary conditions it is
possible to sum interactions of particles with all of their images
residing in cells obtained by translation from the original
computational cell,\cite{lekner,niels} yielding an effective
interaction which is periodic in the computational cell used (see
Fig.\ \ref{fig:lekner}).
On making integral transformations, Coulomb
interactions are computed by summing over fast-convergent Bessel
functions with great accuracy (see Ref.\ 
\onlinecite{niels} for a detailed study of the Lekner summation
technique). 
The computational efficiency is further
enhanced by tabulating the effective interaction. This is
possible since the difference between the obtained effective
interaction and the Coulomb interaction is a well behaved function
which can be easily calculated using polynomial interpolation.

At the beginning of each simulation
we place the holes at random and assign to each hole a magnetic dipole
moment of constant size, but random direction.  We  study the energy
landscape and the dynamics of the system using
 three different methods: Monte Carlo (MC), Langevin molecular dynamics (MD) 
and a hybrid MC-MD method.\cite{janez}  All three
methods yield essentially the same results. 
Both MC and MD methods are well known in the literature\cite{nr}
and hence we only review details pertinent to our calculation.
In the MC method we use the standard Metropolis algorithm for the
acceptance of hole configurations. For the Langevin MD method the
dynamics of the system is determined by the forces,
obtained from Eq.\ (\ref{eq:hamiltonian}), with a noise term,
$\eta_i$, for each particle which satisfies the usual
fluctuation-dissipation condition
$\langle \eta_i \eta_j\rangle = 2\pi \gamma T \delta_{i,j}$, where $i$,
$j$ correspond to  all possible degrees of freedom and $\gamma$ is a
damping term.\cite{nr} Since the
system exhibits several phases (see Fig.\ \ref{fig:phase_diagram}) for some
values of the input parameters, its ground state is not always well defined
and a numerically obtained low temperature state
may, in fact, depend on the initial and boundary
conditions. Hence, in order to rapidly reach a hole configuration with
the lowest global minimum energy we perform simulated annealing from
high temperatures.

The hybrid MC method includes elements of both the MC and MD methods:
the hole configuration is again determined using the standard
Metropolis algorithm, but here a new configuration is obtained by
letting the system evolve through a classical MD calculation
over a certain time period. Note that, in principle, in classical MD
calculations the energy is a conserved quantity, hence every step
should in principle be accepted. In reality the MD method introduces
errors which typically slightly lower the system energy, just as
required by the Metropolis algorithm. Hence
this method yields extremely high MC acceptance ratios.\cite{janez}

\section{results}  % 
\label{sec:results}  % 

In this section we discuss the results of our numerical calculations.
We first present the obtained ordered phases and the phase diagram of
the system and discuss its implications. We then show the hole
ordering in the presence of disorder. Finally we show the properties
of the system as a function of $T$ and in particular the dynamics of
the observed (stripe) pattern formation.

Before we begin with the presentation of our results we address
the input parameters of the model, namely the hole density,
$\sigma_s$ {(or the doping level $n$)}, 
the strength of the isotropic and dipolar part of the AF
interaction, $A$ and $B$ respectively, the AF correlation length,
$\xi$, the temperature $T$ and the concentration of impurities, $c_i$,
for the systems with static point disorder. {We 
define the doping level $n$ as the hole density measured in 
units of cuprate lattice spacing; thus $n=1$\% corresponds 
to 1 hole per 100 $a^2$, where $a\approx 3.8$ \AA.}

The input parameters are not necessarily independent of each other, e.g., $A$
and $B$ should be proportional to each other, with
$A \approx U$ when $t\gg U$ and 
$A \approx 4t^2/U$ when $U\gg t$ (see Ref.\ \onlinecite{david}).
However, since the range of the bond breaking and dipolar interactions
is vastly different, it is reasonable to treat $A$, $B$ and $\xi$ as
independent parameters. Clearly both $A$ and $B$ should be of order of the
Hubbard $U$ in the SDW approach and of the order of $J$ in the strong
coupling limit, and the correlation length is of order 1 to 10 lattice
spacings in cuprates.

We begin with the low temperature properties (ground state) of the
system as a function of $B$. 
The relevant  order parameter for charge ordering
is the Fourier transform of the hole density:
\begin{eqnarray}
\rho({\bf q}) = \frac{1}{N} \sum_{i=1}^N e^{i {\bf q} \cdot {\bf r}_i} \, ,
\label{rho}
\end{eqnarray}
where ${\bf r}_i$ is the position of the $i^{th}$ hole and $N$ is the
total number of holes. A peak in $\rho({\bf q})$ at some wave-vector
${\bf q}={\bf K}$ indicates ordering. Returning to the SDW picture,
presented in Sec.\ \ref{sec:model}, we recall that the hole density is
given by
\begin{equation}
\rho_k = \sum_{q,\alpha} h^\dagger_{k+q,\alpha} h_{q,\alpha}.
\end{equation}
From the definition of 
the staggered magnetization, Eq.\ (\ref{eq:staggered}), it is
immediately clear
that $\langle S_z(q) \rangle \propto \delta_{q,K+Q}$, i.e., a peak in
$\rho_k$ at $K$ leads to a magnetic peak at $Q+K$ and by symmetry at $Q-K$.

\begin{figure}
\centerline{
{\epsfxsize=8.0cm \epsfbox{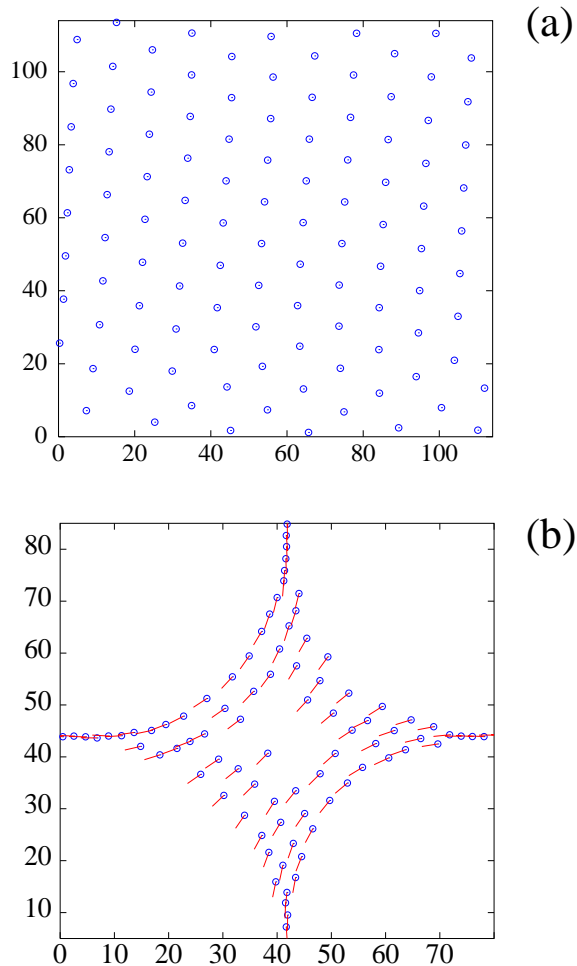}}
}
\caption{Low $T$ states of the hole vacancy system.
(a) For $B=0$ the holes (circles) form the Wigner crystal and (b) for 
$B\rightarrow \infty$ (an unphysical case)
they form a ``clump'' pattern, a characteristic
of the ``mesoscopic phase separation.'' The 
dipole orientation (shown by the
segments, originating from the circle centers)
indicates finite magnetization at each star cluster. {In both panels
the doping level is $n=15$\%.}}
\label{fig:wigner}
\end{figure}

Since the interaction due to the second term in Eq.\ (\ref{total})
(isotropic attractive interaction)
is extremely short-range (in fact in an infinite system it is a $\delta$
function), it is initially 
reasonable to set $A=0$ and explore the behavior of the system as a
function of $B$. We return later to
the role of $A$.  As explained in Sec.\ \ref{sec:model} low
$T$ properties have been obtained by annealing the system from some
high temperature ($T\sim 5000$K) down to temperatures of order 1K.
In the extreme case $B=0$ we find a Wigner crystal with small distortions, 
to be the state of lowest energy, as expected\cite{wigner} 
(see Fig.\ \ref{fig:wigner}a).
The small distortion of the Wigner crystal structure is due to the
periodicity, which introduces a small spatial anisotropy into the system
due to the shape of the computational box.  Indeed, upon setting,
e.g., $L_y/L_x=\sqrt{3}/2$, we obtain a perfect Wigner crystal to be
 the ground
state. Another extreme case is when $B\rightarrow\infty$. In this case
the AF dipolar interaction
dominates over the average Coulomb interaction;
one then finds star shaped clumps of
holes, similar in shape to those found in Ref.\ \onlinecite{gooding-stars},
which can, at sufficiently high density, form a geometric
structure (e.g., a Wigner crystal of clumps). We note that this case
is rather unphysical, as macroscopic phase separation is inconsistent
with our initial assumption of the two-body dipolar interaction being
independent of the many-hole effects.
On the other hand, ordering of macroscopically hole rich regions
 is in agreement with the conjecture that all ground states are
geometrically ordered.\cite{pwa}

On increasing $B>0$, at fixed density, the Wigner crystal becomes 
unstable and a new
phase with diagonal stripes is formed{, as shown in Fig.\ 2a of Ref.\
\onlinecite{sybcj}}.  
The main characteristic of
this phase is a
ferro-magnetic ordering % 
of the AF dipoles.
The situation here is
very similar to that observed in La$_{2-x}$Sr$_x$NiO$_{4+y}$ (see 
Ref.\ \onlinecite{nickel}).  
Note that {such} a state with 
a dipole ordering appeares to violate time reversal symmetry.
On the other hand the true ground state in this case also involves
hole ordering, with holes aligned in stripes either perpendicular or
parallel to the dipole orientation. However, the interstripe distance, in
this case, is close to that between holes within a stripe and hence a
simulation inevitably yields a ``glassy'' state, with many defects.
This is reflected  in the shape
of $\rho_k$, which shows broad peaks (see Fig.\ 2b in Ref.\
\onlinecite{sybcj}), indicating an
average interstripe distance.

\begin{figure} % 
\centerline{
{\epsfxsize=6.50cm \epsfbox{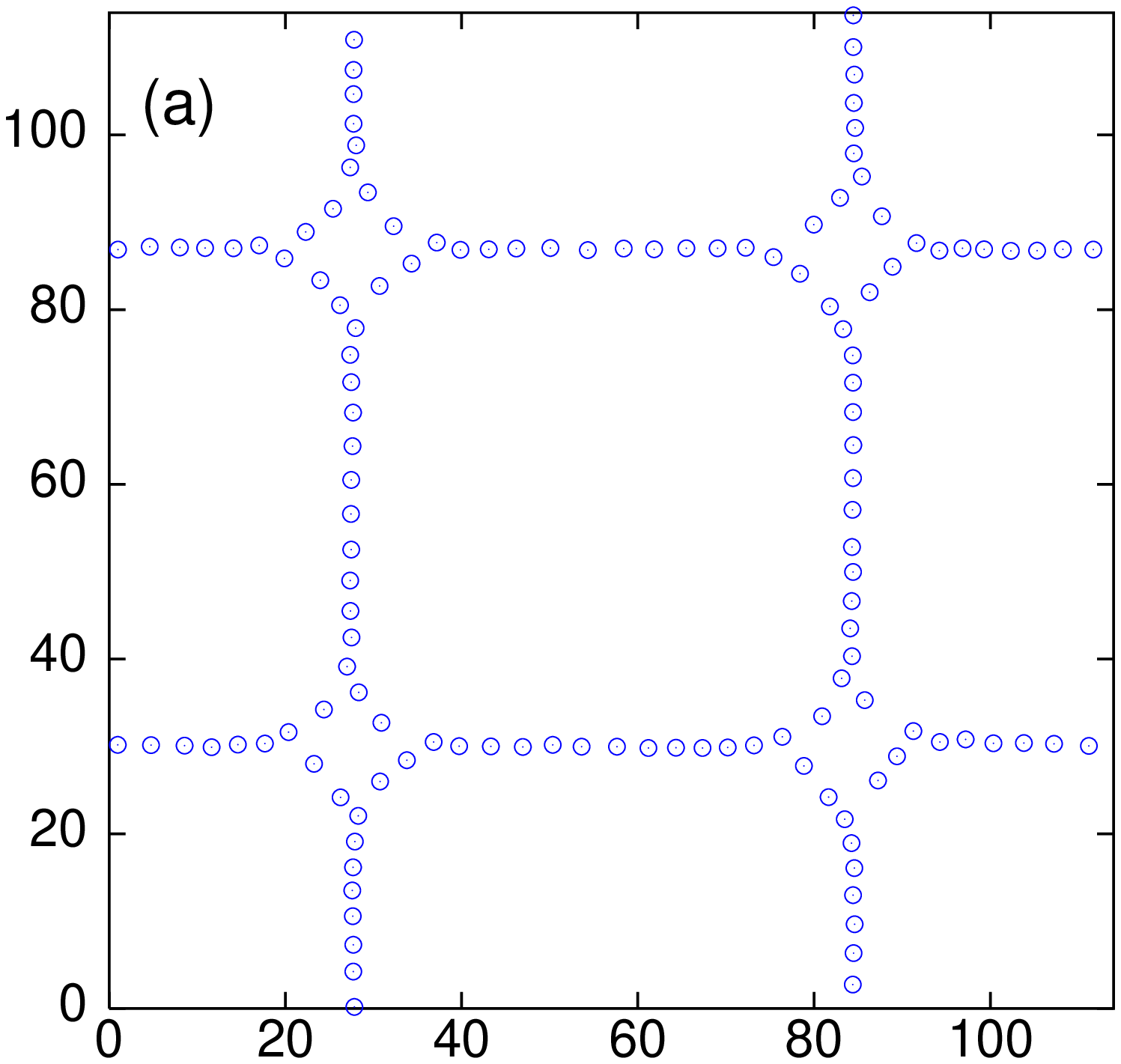}}
}
\centerline{
{\epsfxsize=5.0cm \epsfbox{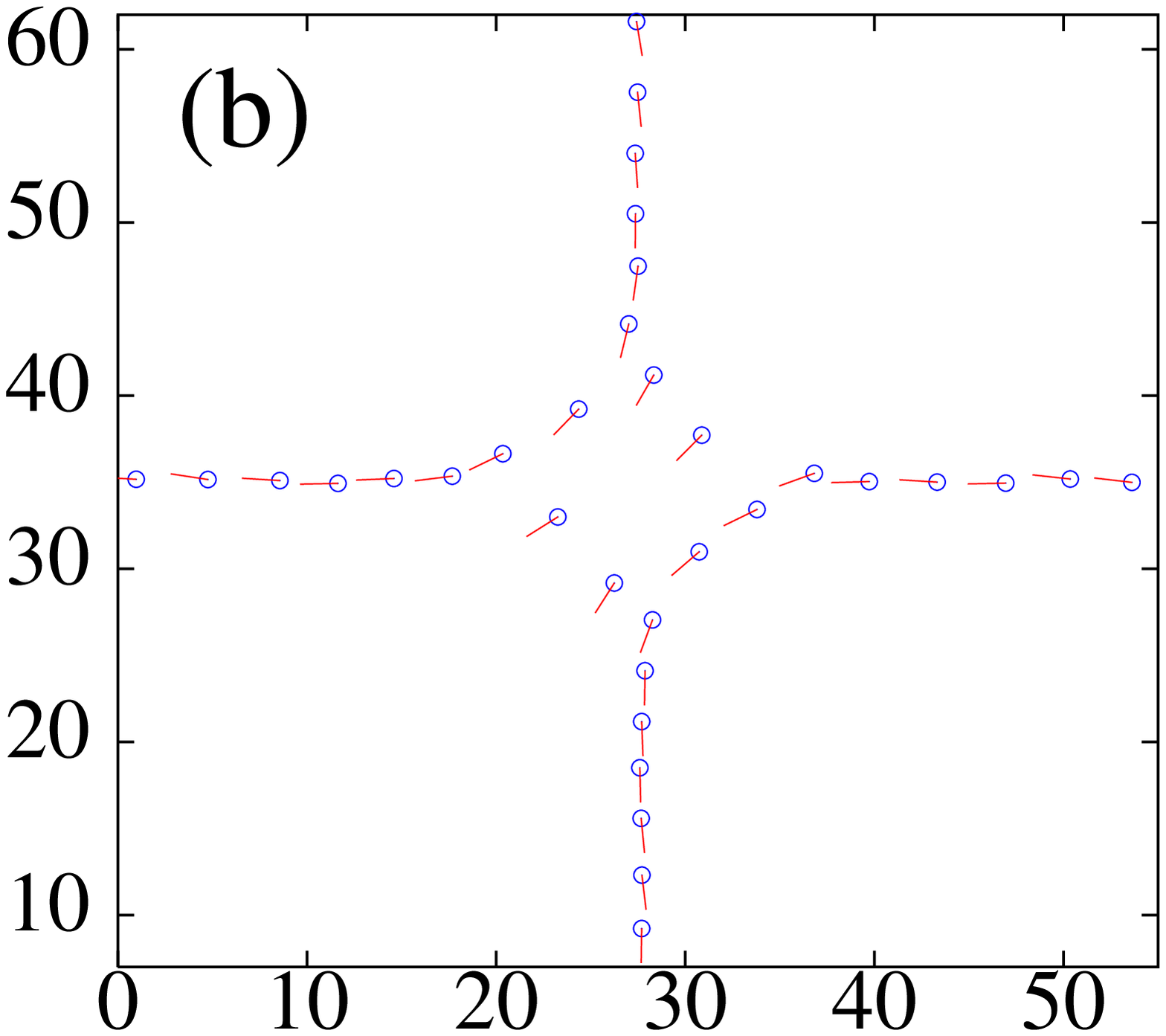}}
}
\centerline{
{\epsfxsize=5.0cm \epsfbox{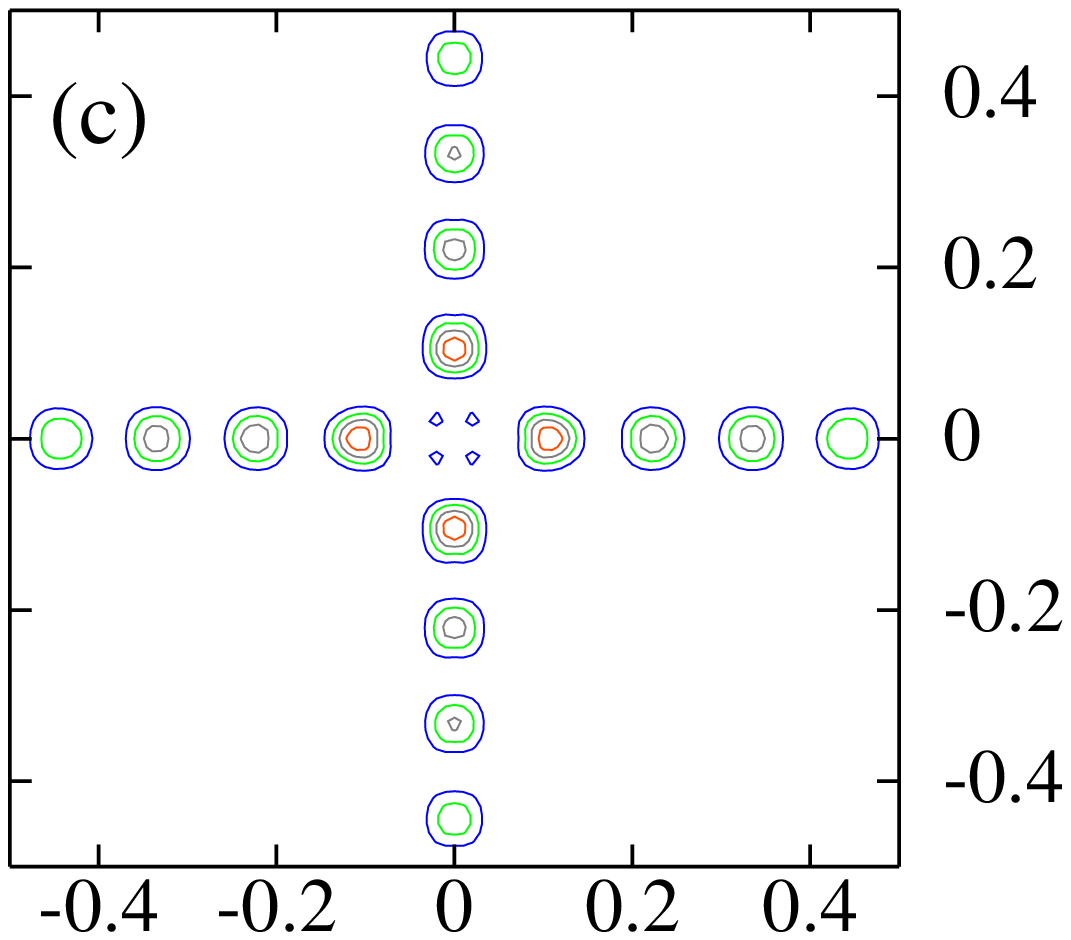}}
}
\caption{Low $T$ state for larger 
  finite values of $B$, $B=4eV${, at $n=15$\% doping level}. 
  The holes (circles) form a grid\cite{sybcj} 
  (panel (a)) with dipoles orientated along line segments and a dipole
  rotation at grid intersections (panel (b)). Panel (c) shows a
  contour plot of the average charge density $\rho_k$ in arbitrary
  units, indicating ``perfect'' geometric order.\cite{sybcj} 
  Even after averaging
  over many solutions in this case the
  charge peaks are much sharper than those 
  {found in the ferro-dipolar phase % 
  (see Fig.\ 2b in Ref.\ \protect\onlinecite{sybcj})}. 
}
\label{fig:tile}
\end{figure}
As shown in Fig.\ \ref{fig:tile}a, at larger values of
$B$ a linear stripe is formed, which, with increasing density tends to
close into a loop.  More importantly, 
the loop formation is accompanied by magnetic dipole orientation
along the straight portion of a loop with gradual rotation by $\pi/2$
at each corner.\cite{vertex} Due to the rotation of dipoles at
corners the loops interact, and eventually form the checkerboard (grid)
pattern.\cite{ferro-vertex}
The size of the distance between holes within
a line is determined by the ratio of $B$ (or the sum of $A$ and $B$,
for $A\neq B$) and the Coulomb energy; the
grid sizes are determined by the hole density alone.  These results
appear to be consistent with the DMRG numerical solution of the $t-J$
model\cite{white} which also finds loops of holes, except that in our
case the periodic boundary conditions and the Coulomb interaction
yield a ``tile grid'' 
as opposed to ``droplets.''  Note the almost perfect
(infinite charge correlation length, $\xi_c$) crystal structure
obtained (Fig.\ \ref{fig:tile}c). 
It is noteworthy that a typical solution yields a finite
dipole moment at each grid intersection, which, in turn, can take one
of the two orientations (along two diagonals of the computational
box), thus creating a highly degenerate system of moments (see Fig.\
\ref{fig:coex} below). 

We recall that the presented solution is obtained assuming 
an arbitrary dipole orientation with a constant hole dipole
magnitude, i.e., a continuum of angles between the dipoles and a fixed 
axis. As explained in the previous section, 
at very low doping the dipoles would reside near the BZ diagonals,
i.e., they would assume almost ``discrete'' orientations. 
In order to study the effect of this ``discreteness'' 
we have performed the same simulation this time assuming that hole
dipoles can take only one of four directions (determined by the momenta of 
the maxima of the lower Hubbard band). We find that the physics of the 
pattern formation is qualitativelly unaltered (hence we do not
present them here), except for one important difference: the
``bending'' of stripes at the grid intersections disappears, i.e., one 
no longer has a finite dipole moment at these intersections.

\begin{figure} % 
\centerline{
{\epsfxsize=7.0cm \epsfbox{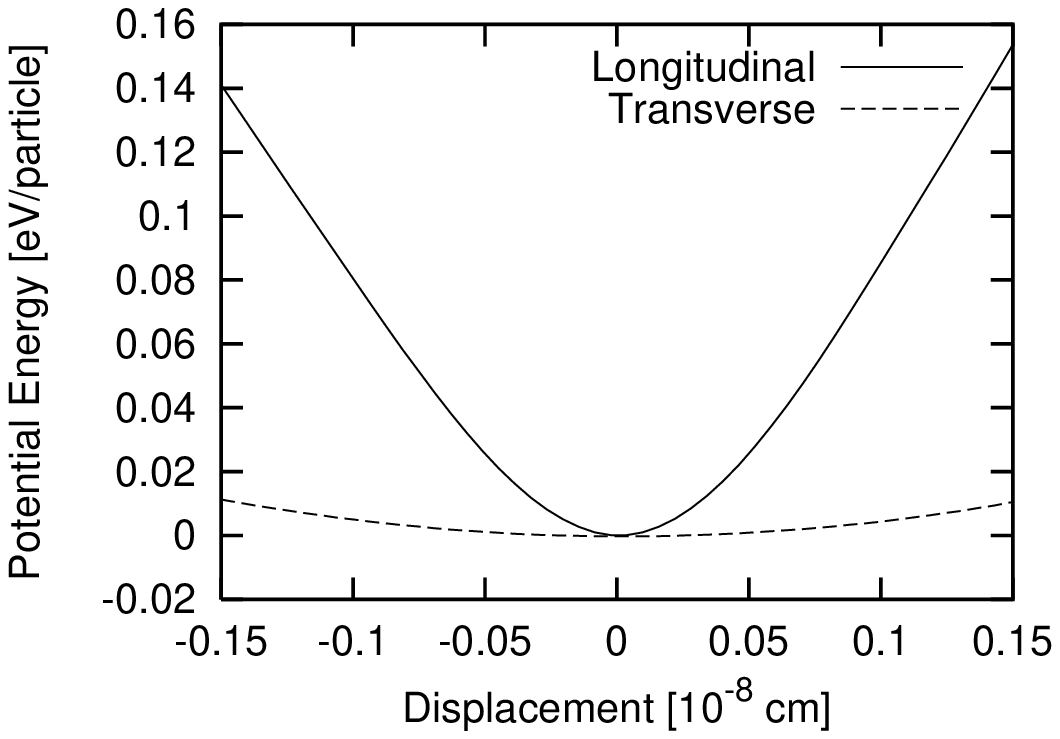}}
}
\centerline{
{\epsfxsize=7.0cm \epsfysize=6cm\epsfbox{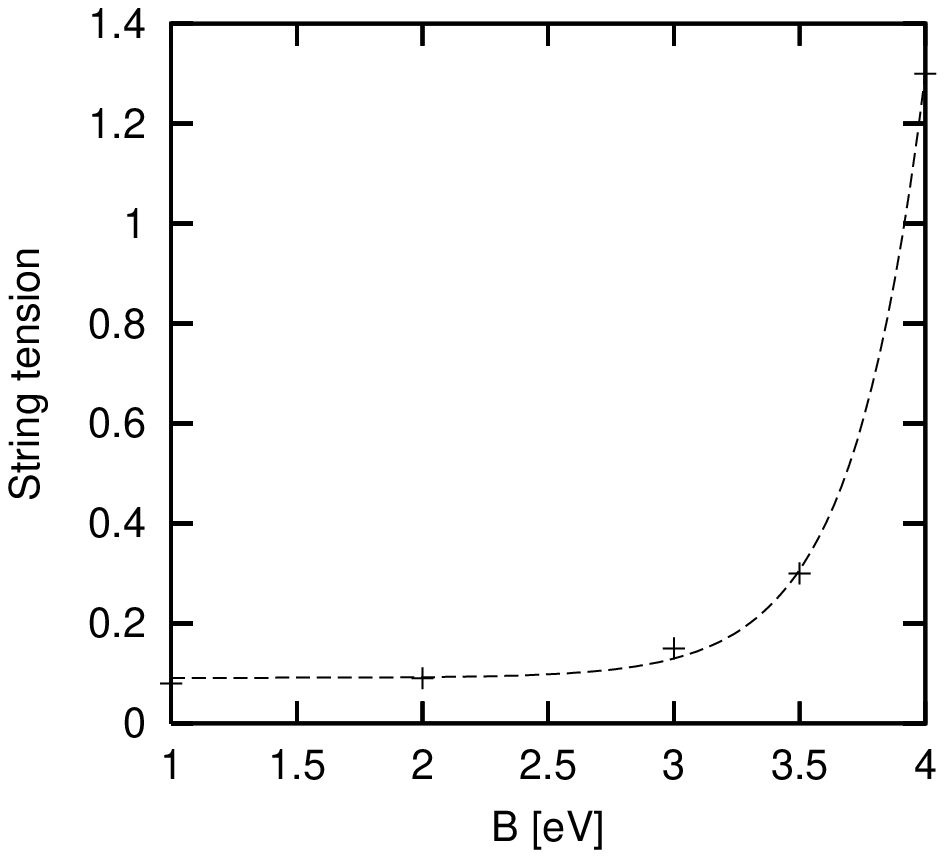}}
}
\caption{Top: Energy as a function of a hole 
position, reflecting the string tension in the stripe, for $B=4$
eV and $n=15$\%. Clearly, the motion of holes perpendicular to the 
the grid directions is quite hard and while the motion along a line 
stripe is much softer (see also text). Bottom: average string
tension as a function of $B$, at $n=15$\% Note the almost exponential 
dependence (solid line)
up to the critical value of $B\sim 4$ eV where the system
undergoes the first order transition between the ferro-dipolar and
grid phases.}
\label{fig:tension}
\end{figure}

Another way of quantifying this ordered phase is by straightforwardly 
calculating the
``string tension,'' which, at $T\rightarrow 0$ is equal to $\partial^2
U/\partial x^2$, where $U$ is the total potential energy {and $x$ is a
small hole (or stripe) displacement}; a large
string tension indicates a high stability of the obtained phase, and
vice versa. In Fig.\
\ref{fig:tension}a we show the string tension for motion perpendicular
to a grid side compared with the motion along a side.
As seen in the figure, the grid phase
(and, as discussed below, the stripe phase) is extremely stable
with respect to the hole motion perpendicular to the holes line
segments, due to the Coulomb interacton. On the other hand, at larger
doping values and fixed hole density the stripes are almost
compressible, i.e., the motion of holes along a stripe is rather
soft. The anisotropy of the perpendicular and longitudinal string
tension decreases with decreasing $B$.

\begin{figure} % 
\vskip1cm
\centerline{
{\epsfxsize=6.50cm \epsfbox{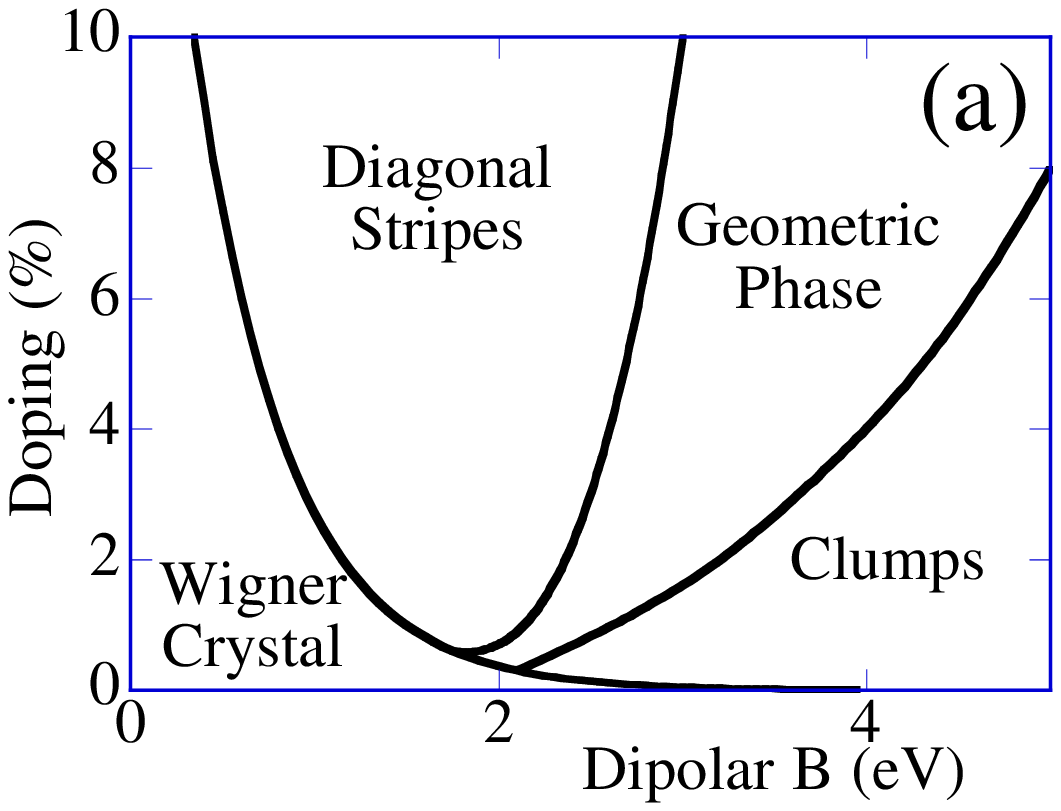}}
}
\vskip1cm
\centerline{
{\epsfxsize=6.0cm \epsfbox{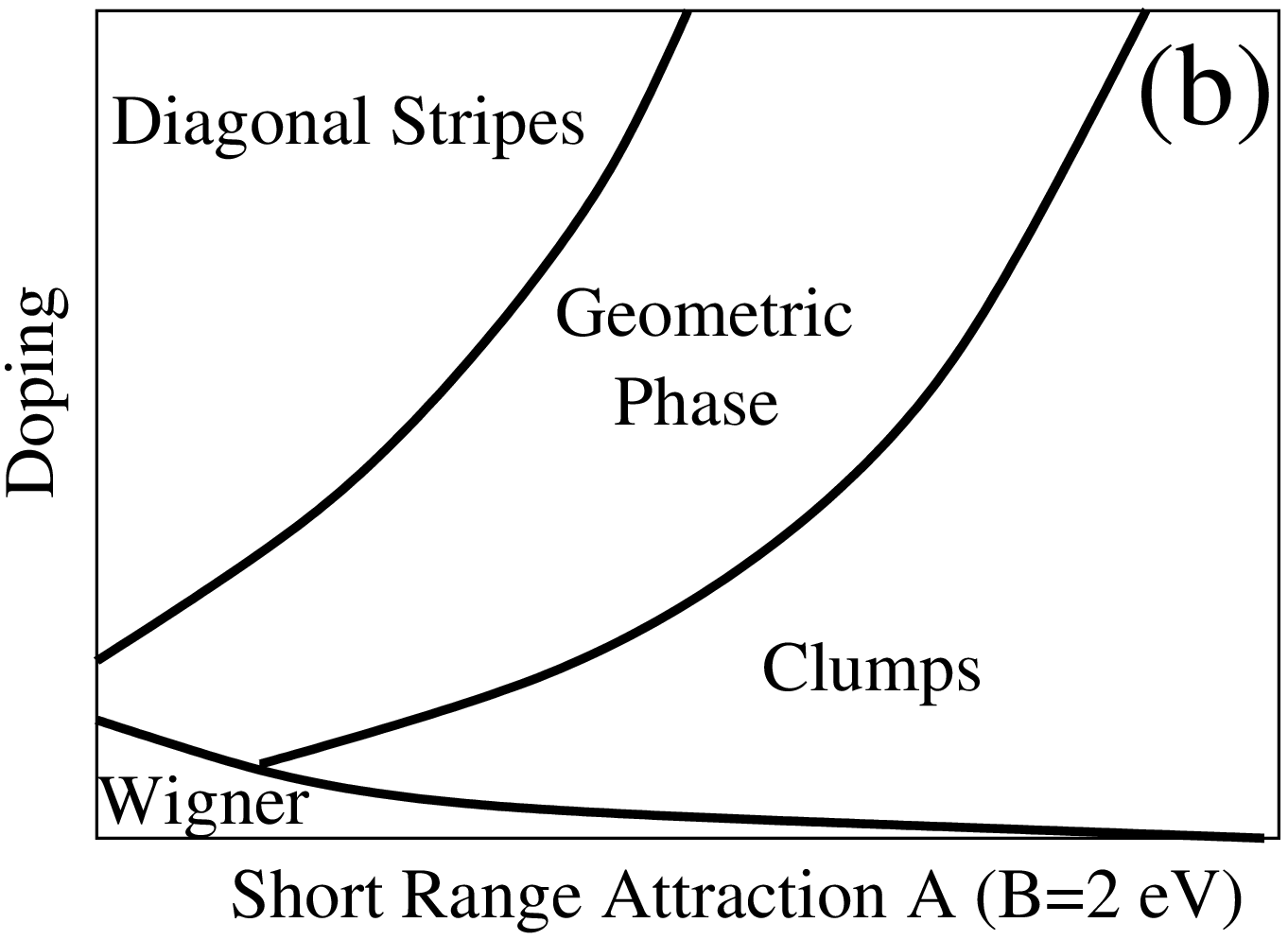}}
}
\caption{(a) $A=0$, $n-B$ phase diagram {(from Ref.\
  onlinecite{sybcj})}.  (b) Fixed $B$, $n-A$ phase diagram. {It 
  is noteworthy that the experimentally relevant values of $A$ and $B$
  are of order $\sim 1$ eV. The slightly higher values of $B$, at which
  we find geometric (grid and stripe) phases, are due to the fact that
  we consider unscreened Coulomb interaction and in reality they would 
  be considerably smaller (see also Fig.\
  \ref{fig:maze})}. 
}
\label{fig:phase_diagram}
\end{figure}

At fixed AF correlation length the four observed phases yield a 
diagram which we show in Fig.\ \ref{fig:phase_diagram}a. 
We remark that in all phases a non-vanishing value of $A$
leads to a decrease in the effective value of $B$ at which the
transitions occur, as shown in Fig.\ \ref{fig:phase_diagram}b.
The isotropic term $A$ alone
{\em never} produces any non-trivial geometric phase (e.g., stripes),
even with inclusion of lattice effects.  We find that the transition
between the ferro-dipolar and the grid phases is first order, as
indicated 
by the coexistence of phases in Fig.\ \ref{fig:coex}. {Note that this
transition always occurs on increasing the doping level to
sufficiently (and artificially)
high values, where our theory need not apply.} No 
coexistence of phases has been observed at
other transitions, suggesting that they are second order. 
{We also recall that our calculations are \emph{quasiclassical}
  and thus the obtained geometric (stripe) phases are
  insulating. Moreover, in our formalism the  
  hole density within a stripe can assume an arbitrary value,
  depending on the dipolar interaction strength.}
\begin{figure}
\centerline{
{\epsfxsize=8.0cm \epsfbox{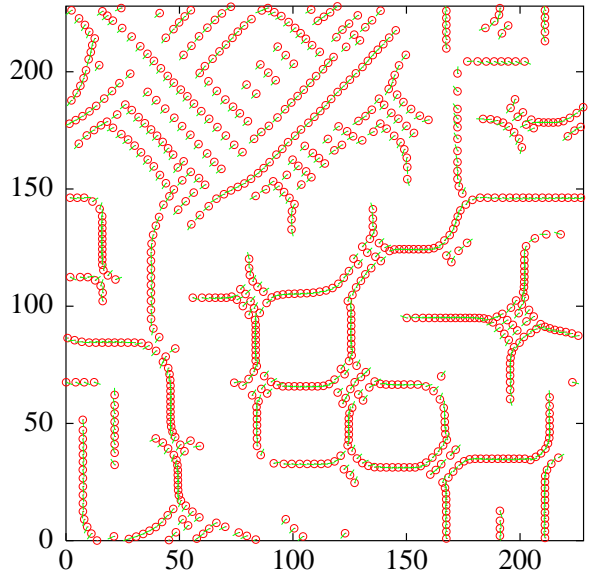}}
}
\caption{Coexistence of dipolar and grid phases, indicating the first
  order nature of the transition between them.}
\label{fig:coex}
\end{figure}

In the cases presented above we have assumed a uniform magnetic dipolar
interaction.  It is well known that there are orthorhombic and
tetragonal distortions in practically all transition metal oxides.
In particular static stripe formation has only been observed in the
low temperature tetragonal phase of
La$_{1.6-x}$Nd$_{0.4}$Sr$_x$CuO$_{4}$\cite{tranquada}.  In order to
study the influence of the anisotropy we assume that the magnetic dipole sizes
along the $x$ and $y$ directions have anisotropy $\alpha$ ($\alpha=1$
corresponds to the isotropic case). Figure
\ref{fig:aniso} shows the pattern we obtain for $\alpha=0.8$:
\begin{figure}
\centerline{
{\epsfxsize=7.0cm \epsfbox{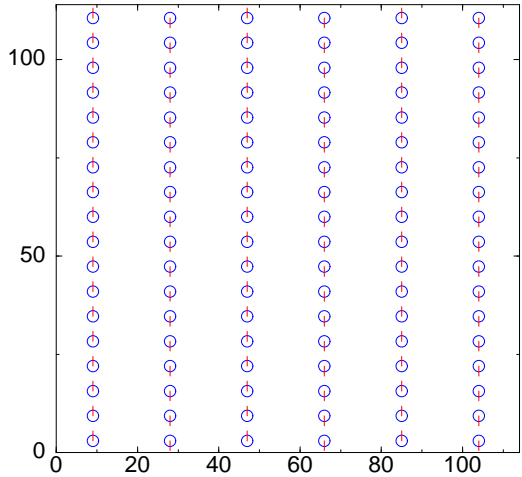}}
}
\caption{With a small
  $x-y$ directional anisotropy in the dipolar interaction, Eq.\ 
  (\ref{total}), with the anisotropy parameter $\alpha=0.8$ (in the
  uniform case $\alpha=1$), the holes (shown by circles) form a
  stripe, rather than a grid pattern. Note the hole dipole
  orientations (shown by the line segments), altering direction in
  neighbouring stripes, corresponding to the $\pi$ phase shift of the
  local magnetization between magnetic domains separated by
  stripes.\cite{emery1}}
\label{fig:aniso}
\end{figure}
\noindent The rotational 
symmetry is broken and a stripe superlattice is formed,
with a charge ordering vector ${\bf K} = (2\pi/\ell) {\bf x}$, where
$\ell$ is the inter-stripe distance.  
More importantly, the total dipole moment in this state vanishes.
As explained in Sec.\ \ref{sec:model}, this yields a Fourier
transform of the magnetization $S({\bf q}) = \langle S_z({\bf q})
\rangle$ peaked at ${\bf Q} \pm {\bf K}$ in
momentum space. Of course, it is reasonable to assume that in
twinned single crystals, used in inelastic neutron scattering
experiments, one has domains which
average out this anisotropy. {Note that in both calculated
geometric phases (see Figs.\ \ref{fig:tile} and \ref{fig:aniso}), the
interstripe distance is much larger than the intrastripe distance, in
agreement with experimental findings in underdoped cuprates.}

\begin{figure}
\centerline{
\epsfxsize=7.0cm \epsfbox{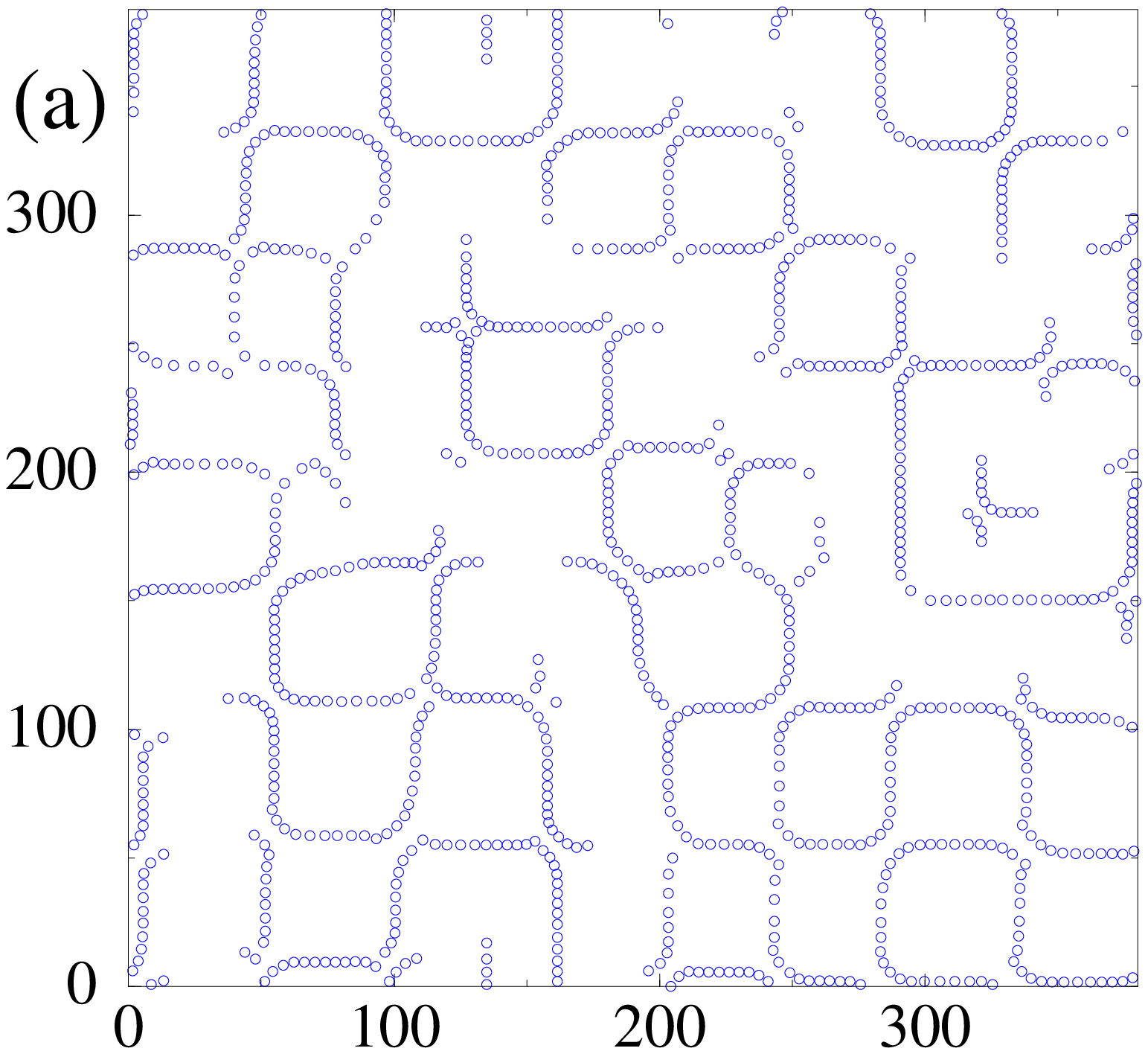}
}
\vskip-1.5cm
\centerline{
\epsfxsize=9.0cm \epsfbox{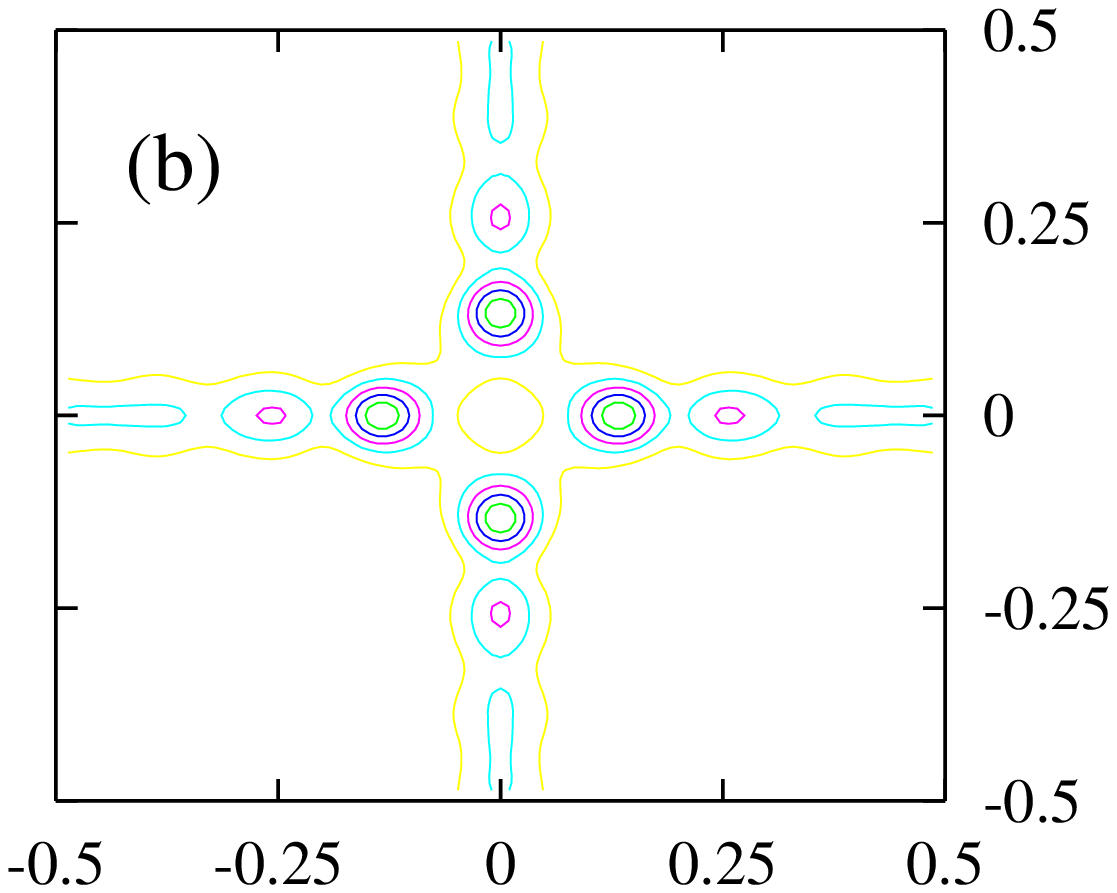}
}
\vskip-1.5cm
\caption{(a) A typical low energy state in a 
larger computational box, obtained for the same value of $B$ and $n$
as in Fig.\ \ref{fig:tile}. Note the presence of point and line
defects. It is important to realize that the average size of the grid
units depends only on the physical parameters and very weakly on the
size of the computational box.
(b) The charge density, $\rho_k$, averaged over a number
(of order 10) of hole configurations. Note that the peak positions are 
the same as those shown in Fig.\ \ref{fig:tile}, but due to the
presence of defects the intensity of the higher Bragg peaks vanishes.}
\label{fig:defects}
\end{figure}

\begin{figure}
\centerline{
\epsfxsize=7.0cm \epsfbox{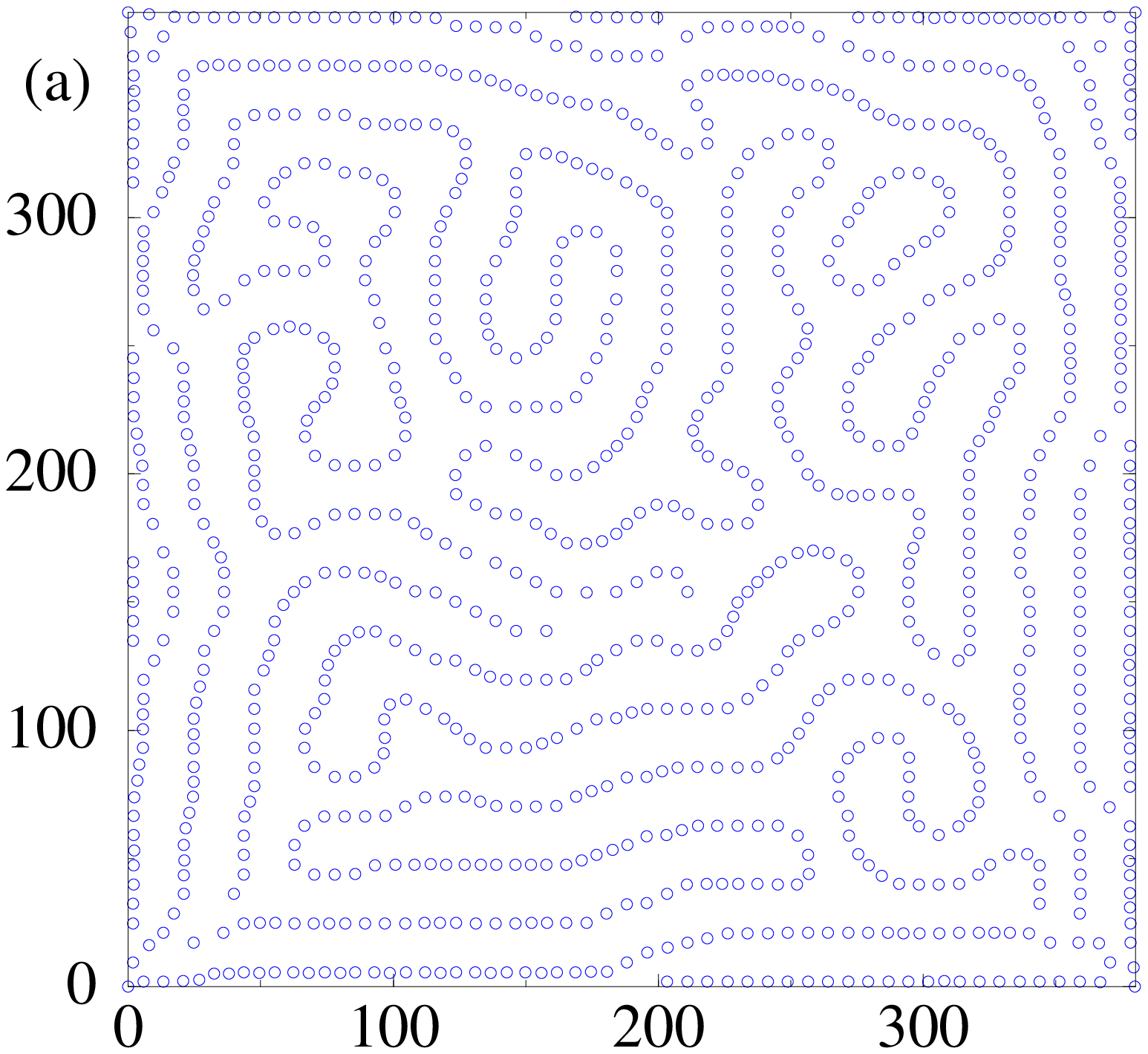}
}
\centerline{
\epsfxsize=7.0cm \epsfbox{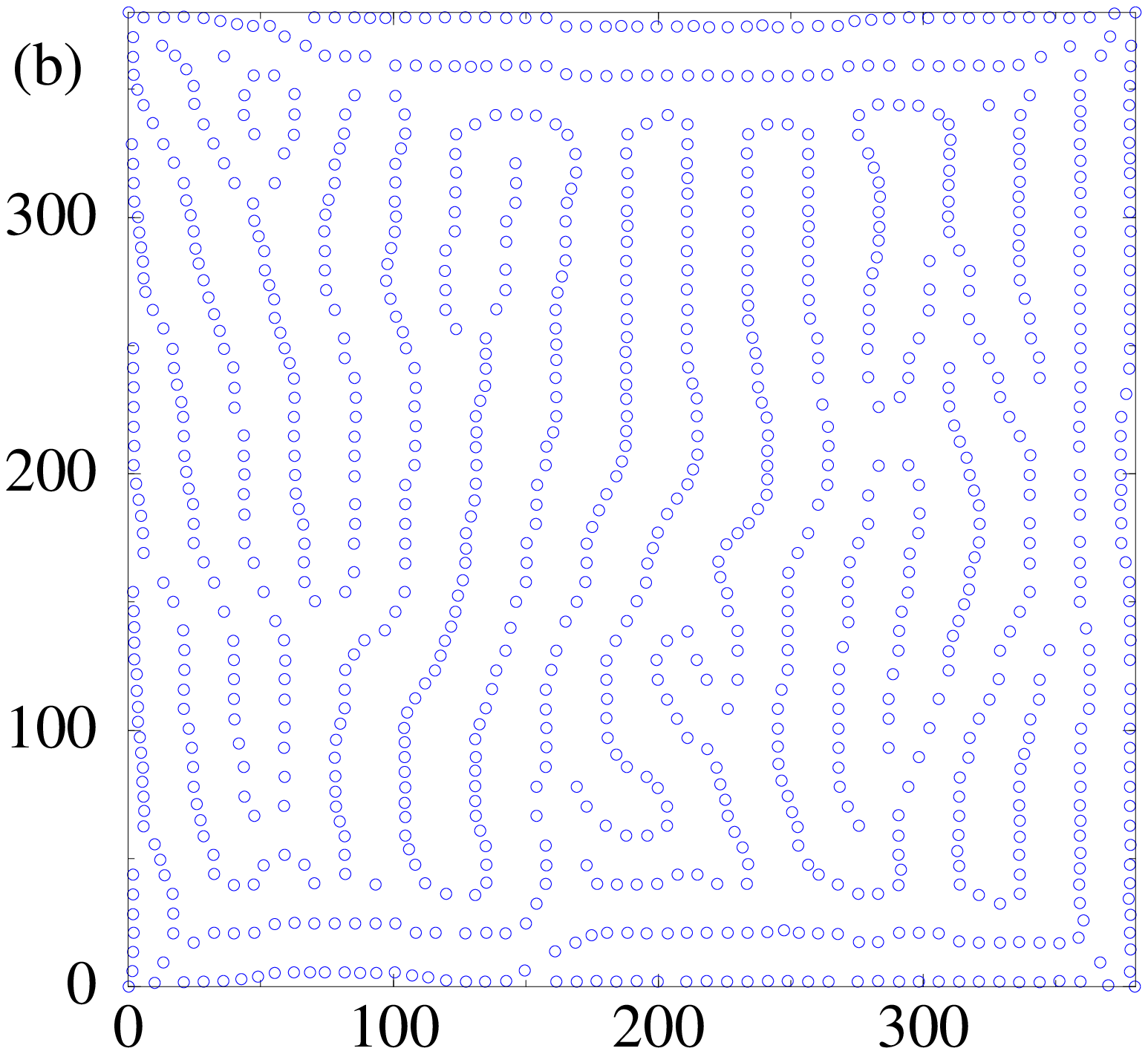}
}
\centerline{
{\epsfxsize=4.0cm \epsfysize=4.0cm \epsfbox{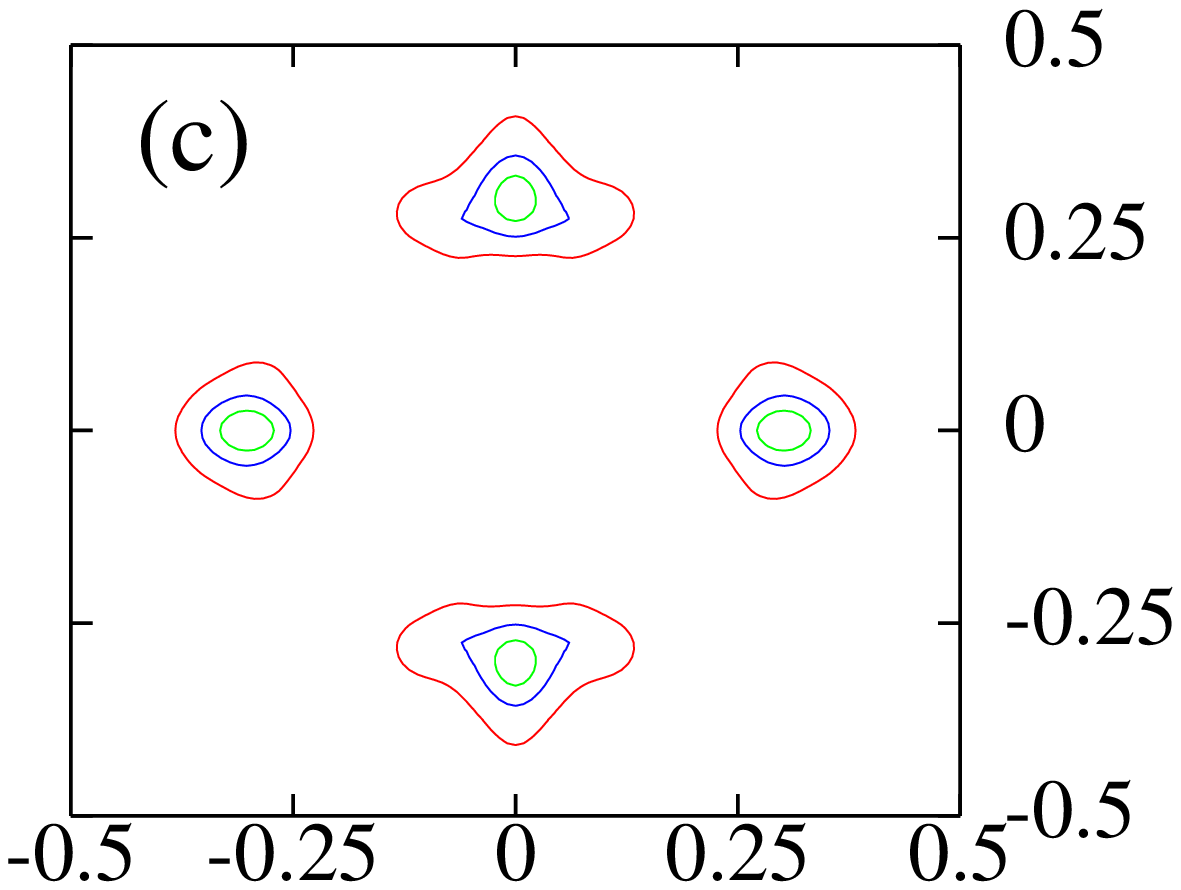}}
{\epsfxsize=4.0cm \epsfysize=4.0cm \epsfbox{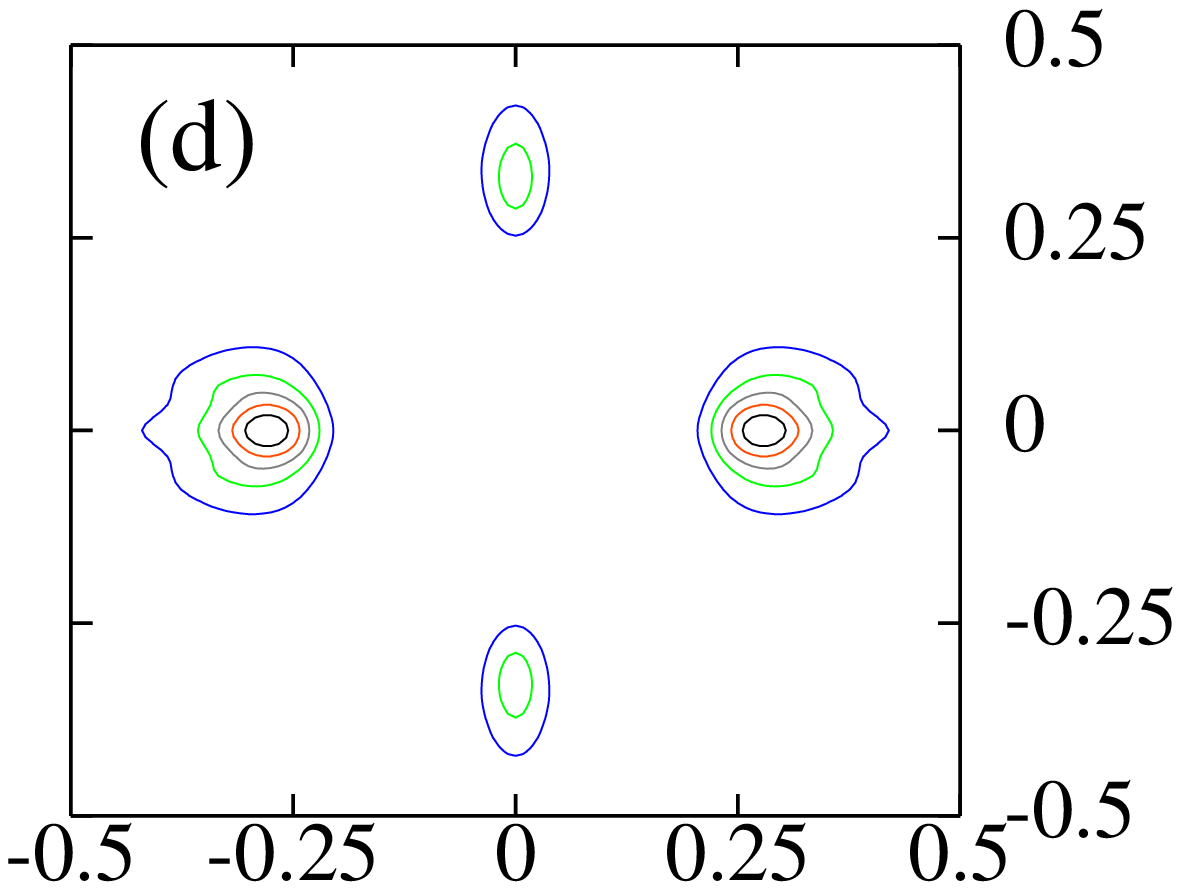}}
}
\caption{(a) A typical low $T$ metastable state of the hole system, obtained
  \emph{without} periodic boundary conditions and with an appropriate
  positive background included. While the ground state is still a
  geometrically ordered state, it is practically unreachable due to
  the presence of many metastable states. 
(b) Adding very small directional
  anisotropy ($\alpha=0.9$, with $\alpha=1$ corresponding to the
  uniform case shown in panel (a); see also Fig.\ \ref{fig:aniso})
  yields stripes with some defects. Panels (c) and (d) show
  $\rho_k$ as a function of momentum corresponding to the results
  shown in panels (a) and (b), respectively.
}
\label{fig:maze}
\end{figure}

Our results are somewhat sensitive to the applied boundary
conditions. 
First, for a small computational box 
the exact size of the grid depends on its
commensuration with the box length, which, in turn depends on
the density. On increasing of the size of the computational box, the
grid size depends only on the physical parameters, as explained below
Fig.\ \ref{fig:tile}. In addition, for a large computaitonal box the
grid pattern, shown in Fig.\ \ref{fig:tile}a, acquires point or
line defects, shown in Fig.\ \ref{fig:defects}a. 
This leads to the reduction in the higher
order peaks observed in Fig.\ \ref{fig:tile}b with no change in their
wave-numbers, indicating the finite value of $\xi_c$, as seen in Fig.\
\ref{fig:defects}b.
The sensitivity to boundary conditions is further seen in 
finite size calculations, i.e, not assuming periodic boundary
conditions, but with an appropriate neutralizing
charge background. In this case 
the holes do not form geometric phases, although
they still form stripes, as seen in Fig.\ \ref{fig:maze}a. 
However, in a finite system even
very small anisotropy ($\alpha\sim 0.95$) again leads to stripe formation,
as seen in Fig.\ \ref{fig:maze}b. It is worth mentioning that the
stripe formation occurs for much smaller values of $B$ and the same
density and AF correlation length in a finite system.

We have also performed simulations in the
presence of a realistic underlying periodic lattice and have found
that this creates slight distortions in the phases, pinning
loops more strongly.  In
particular, the peaks in $\rho({\bf q})$ sharpen in some of these
phases.

\subsection{Role of disorder}

We now turn to the effects of point disorder. There are several kinds of
impurities which are of experimental 
interest in transition metal oxides. We
divide them into four distinct groups: 
\begin{figure}
\hbox{}\hskip-2mm{\epsfxsize=4.1cm\epsfysize=3.8cm
\epsfbox{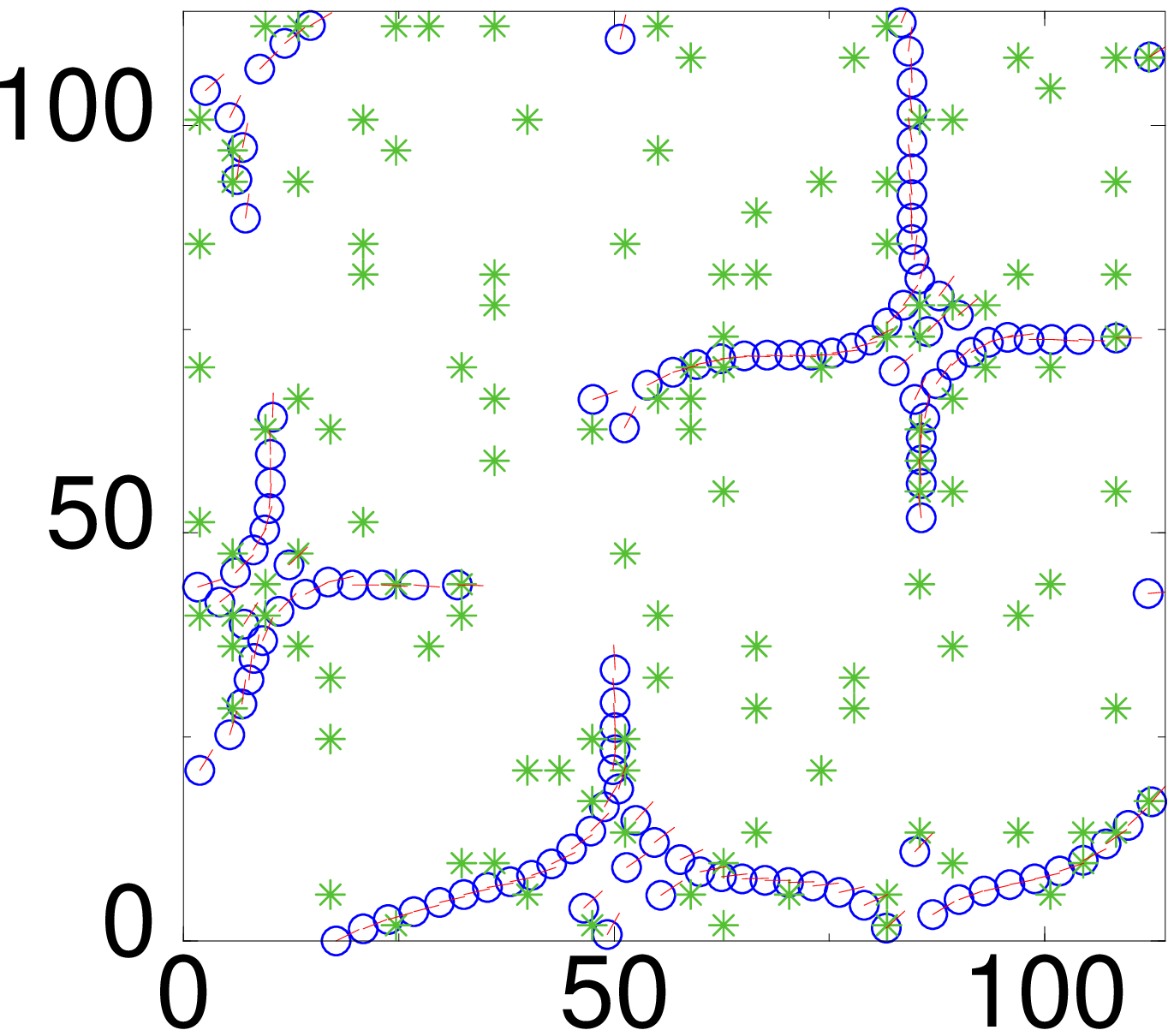}}\vskip-4.3cm
\hbox{}\hskip4.1cm{\epsfxsize=4.2cm\epsfysize=3.9cm
\epsfbox{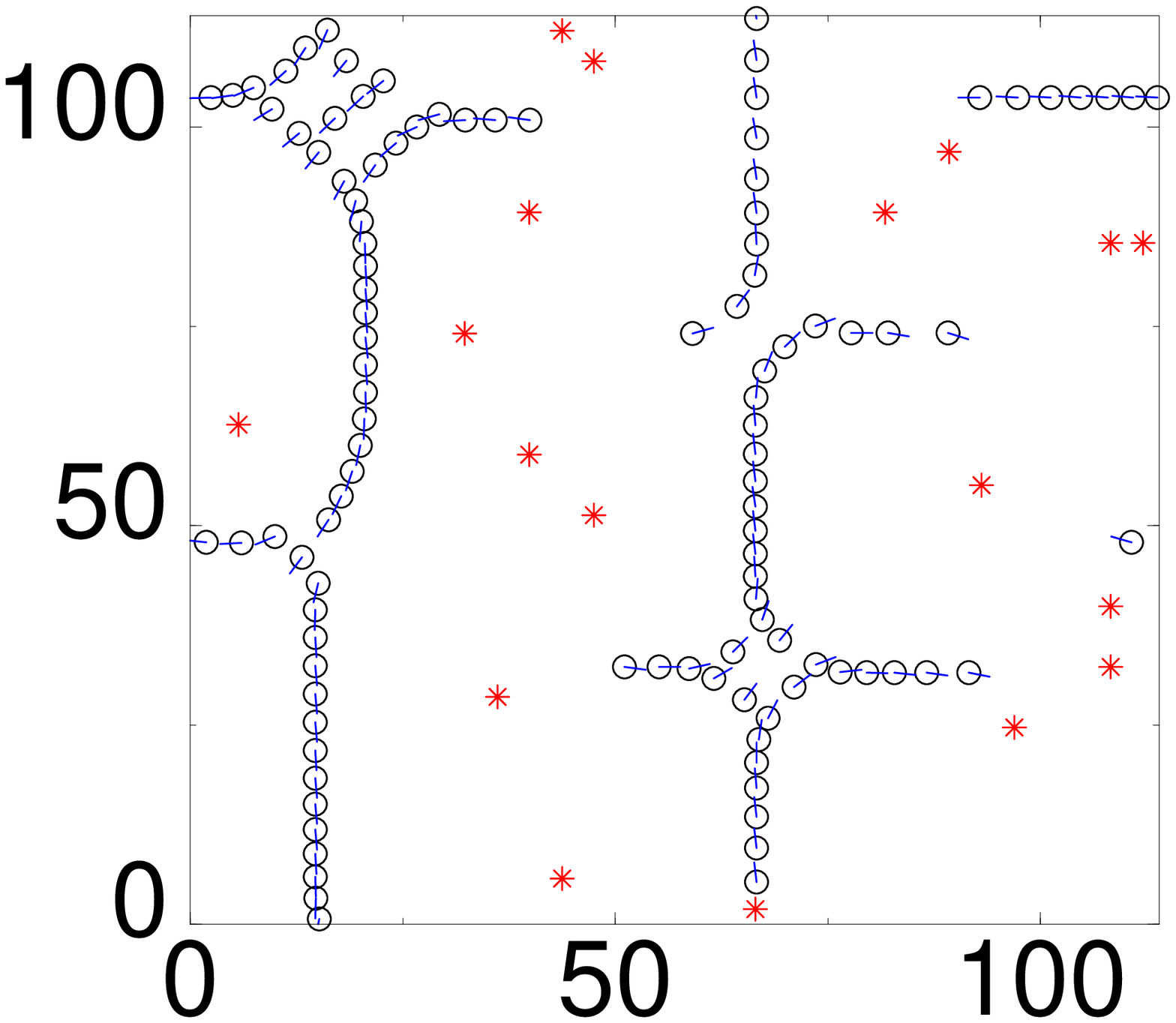}}% 
\vskip.1cm\hbox{
{\epsfxsize=4.2cm \epsfysize=3.8cm \epsfbox{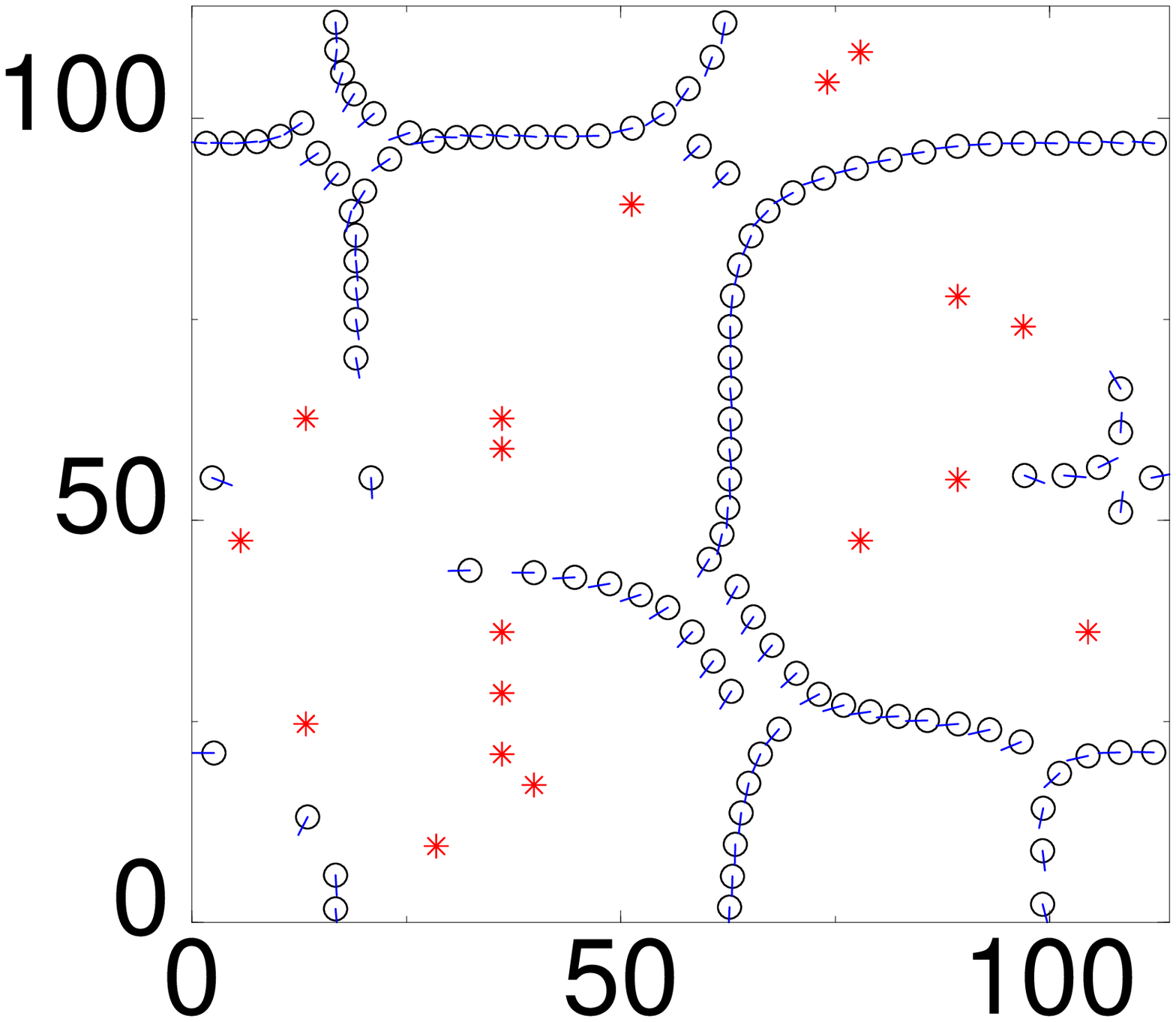}}
{\epsfxsize=4.2cm \epsfysize=3.8cm \epsfbox{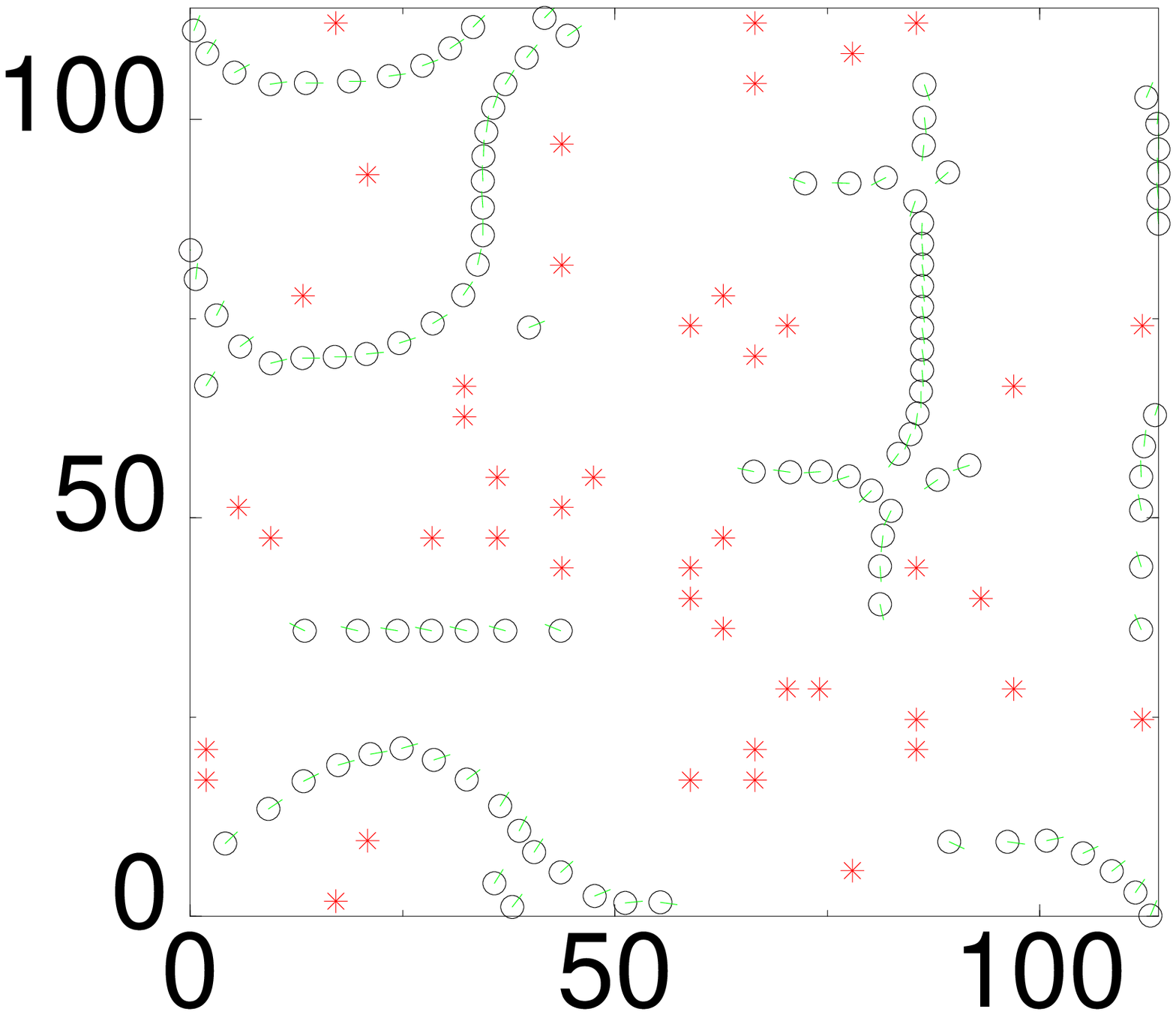}}}
\caption{The effects of impurities of different types. The four panels 
  show hole
  (circles) and impurity (stars) positions, for the case of:
charged impurities, placed $d=6$\AA out-of-plane (top left), 
in-plane repulsive uncharged impurities (top right), in-plane repulsive charged
impurities (bottom left), and in-plane impurities with a local magnetic 
moment which destroys local AF ordering (bottom right). Clearly, all
  impurities destroy the stripe order, although the
  out-of-plane impurities  and the uncharged in-plane impurities are
  nearly as effective as the
  in-plane charged impurities. The latter lead to the a glassy phase
  of stripe segments at relatively low concentrations, of order a few 
  percent.}
\label{fig:4impurities}
\end{figure}
\noindent (I) out-of-plane impurities, such as Sr in LSCO compounds,
(II) in-plane charged impurities, such as (presumably) 
Li,\cite{charged_ion} (III)
in-plane uncharged impurities, such as Ni and 
(IV) in plane uncharged impurities which induce a magnetic moment,
such as Zn.  Hence, we model the impurity effects by
adding a random (in position) potential to the Hamiltonian
(\ref{eq:hamiltonian}), which is either short-ranged and located in plane
or, as in the case of charged impurities, long-ranged (Coulomb) and
either in-plane or out-of-plane (a distance $d$ from the plane where
the holes are located). In the case of type IV we have also
altered the dipolar interaction in the vicinity of an impurity, i.e.,
the magnetic interaction is multiplied by a factor $\tanh(r/R_i)$,
where $r$ is the distance of a hole to a nearby impurity and $R_i$ is
the effective radius of the impurity, which, for the case of Zn, has
been estimated to be of order 2 lattice sites around each impurity
atom.\cite{monod} 
Examples of the effects of the four types of
impurities are presented in Fig.\ \ref{fig:4impurities}.

Clearly, all four types of impurities lead to the destruction of the
geometric (stripe or grid) hole order at sufficiently large impurity
concentration, $c_i$. On the other hand, in all four cases stripe ordering
persists through the formation of line segments of holes, 
resulting in a new, \emph{glassy} phase.\cite{gooding-glassy} Moreover, 
the four impurity types exhibit different mechanisms for destroying the
stripe order. The charge ordered phases are practically unaffected by
a small concentration of uncharged impurities (see Fig.\
\ref{fig:4impurities}b), i.e., the stripes simply avoid impurity
sites. Consequently, the stripes persist to relatively high
concentrations of this type of disorder.

The charged impurities first lead to stripe deformation,
i.e., the stripes 
pass either very close to the impurities (for attractive, pinning 
impurities)
or very far from the impurities (for repulsive impurities), in order to
maximize the potential energy (see Figs.\ \ref{fig:4impurities}a and
\ref{fig:4impurities}c).  With increasing $c_i$ the stripes
rupture and only stripe segments persist. Finally, the impurities
with a local magnetic moment affect the formation of the spiral spin
phase, responsible for the (attractive) dipolar interaction. Since the
magnetic interaction is strongly suppressed in the vicinity of such an
impurity site, even the stripe segments cannot exist there, as
shown in Fig.\
\ref{fig:4impurities}d. 

Impurities are especially effective in destroying the
ordered phases found at small $B$. 
For example, the Wigner crystal state becomes glassy
at relatively low impurity concentrations.  This happens because,
e.g., in the case of impurity type I, the
attractive Coulomb energy between impurities and holes scales like
$e^2/d$, where $d$ is the distance between the planes in which the
impurities and holes reside, 
while the average inter-hole Coulomb energy behaves like $e^2
\sqrt{\sigma_s}$. Thus when $\sigma_s<1/d^2$ the holes are pinned by
impurities.

\begin{figure}
\centerline{
{\epsfxsize=7cm \epsfbox{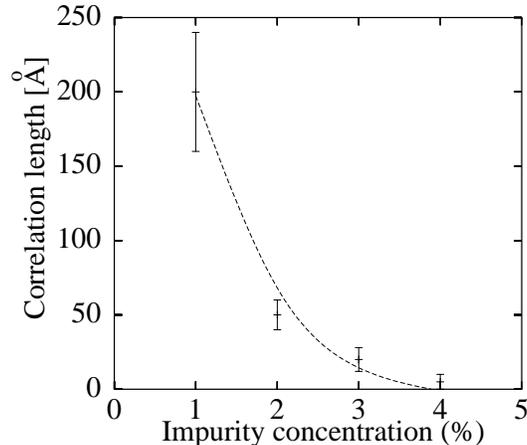}}
}
\caption{Zero temperature {\em charge} 
correlation length as a function of impurity concentration for the
in-plane charged impurity case: the result
is obtained by measuring
the static correlation length, and then averaging over many (of
order 20) different impurity configurations.}
\label{fig:imp_correlation}
\end{figure}

In general the role of impurities depends strongly on the impurity
concentration, $c_i$.
  However, the magnetic dipole interaction is
sufficient to retain the main orientation, as seen in Fig.\
\ref{fig:imp_correlation} where we have plotted the correlation length
as a function of $c_i$. 
This leads us to
conjecture that with the addition of the kinetic energy the holes can
move in string segments in an
orientation given basically by the phase diagram of the clean system.
The stripe motion would then be caused by
mesoscopic thermal or quantum tunneling of the finite strings between
the minima of the overall potential. This would lead to
non-linear field dependence in the low temperature 
conductivity.\cite{bardeen} 

\subsection{Finite temerature results}
We now proceed to the finite $T$ results. The numerical procedure
is identical, except that the temperature is lowered adiabatically to
a finite value (i.e., a numerical annealing). 
In a classical simulation this is equivalent to
introducing kinetic energy into the system. 

\begin{figure}
\centerline{
{\epsfxsize=6.3cm \epsfysize=6cm \epsfbox{wiggly2.bps}}
}
\centerline{
{{\epsfxsize=6.3cm \epsfysize=6cm \epsfbox{wiggly3.bps}}}
}
\centerline{
{{\epsfxsize=6.3cm \epsfysize=6cm \epsfbox{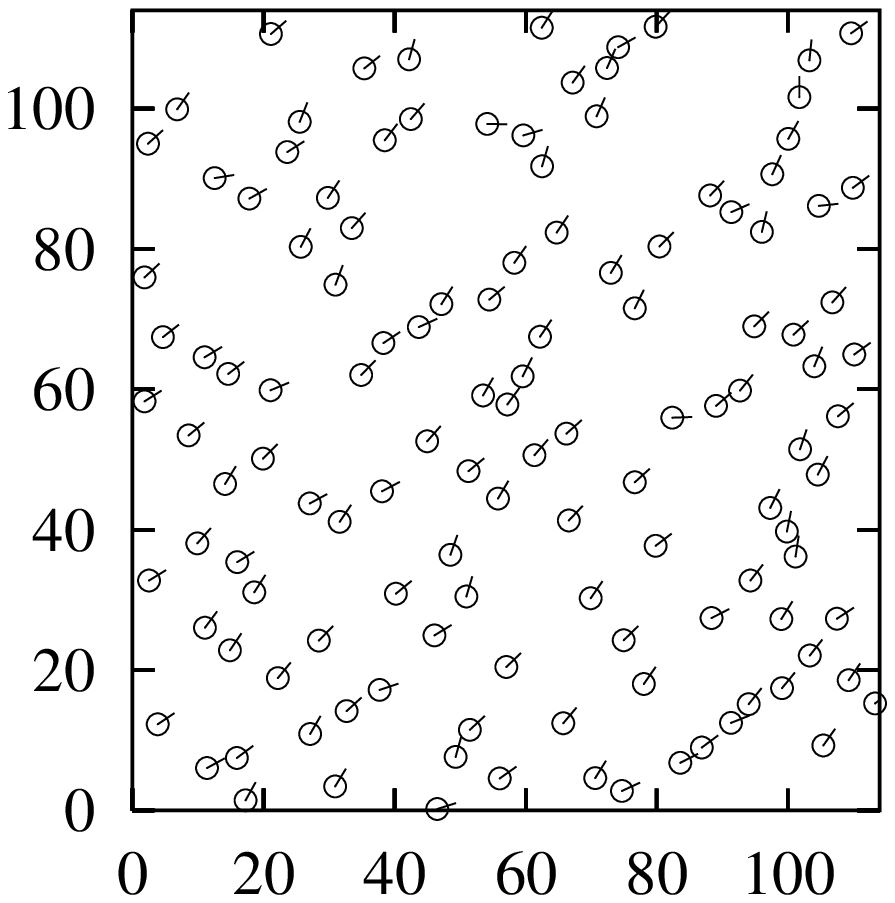}}}
}
\caption{Spatio-temporal 
intermittent behavior of hole ordering \emph{near the melting transition}:
three snapshots are shown. Panel (a) corresponds to a state which is
clearly withing the basin of attraction of the ground state (the same
pattern, albeit deformed), while panel (b) shows a state which is
within the basin of attraction of another low lying geometric state
with more dense stripes along one of the axis. Panel (c)
shows the melted \emph{nematic crystal}--like phase with the hole
dipole moments aligned.}
\label{fig:intermittent}
\end{figure}

In the case of a Wigner
crystal, $B=0$, we find that the introduction of finite $T$ melts the
crystalline structure and the resulting phase is the 2D Coulomb gas.
The diagonal (glassy) phase is also unstable at relatively low
temperatures. On the other hand, the geometricaly ordered states, 
for $a_s\ll\xi$, where $a_s$ is the
distance between holes within a stripe segment,
are all stable up to $T$ of order $\sim B/\sigma_s a_s^2$. At even
higher temperatures, the stripe array melts with a temporal
intermittency of the observed pattern: i.e., spatio-temporal intermittency. 
Fig.\ \ref{fig:intermittent}
shows four stages of this melting process. 
We observed that the stripe melts through a rupture which results in
creation of finite stripe segments that eventually (at constant and
high $T$) disperse into individual holes.

Note that the temporal
geometric pattern (panel (b)) is not the same as that in the ground
state. As mentioned before, there are many low lying geometric states,
close in energy to the ground state, which can temporarily occur at
finite $T$. Hence the dynamics of the stripe ordering is similar to
that observed 
in glasses, characterized by non-gaussian fluctuations.
To show this we follow the dynamics of the hole system at temperatures 
slightly \emph{below} the melting temperature: we start from a
low lying metastable state, such as that depicted in Fig.\
\ref{fig:intermittent}b, increase the temperature adiabatically to the 
point at which the structure begins to melt (which is a measure of the 
activation energy) and let the system
equillibrate. 
\noindent In Fig.\ \ref{fig:spectrum}a we plot 
an energy histogram at this temperature (with the energy shifted by an 
arbitrary additive constant), thus indicating an intensity
of the energy states (bands):
\begin{figure}
\centerline{
{\epsfxsize=6.50cm \epsfysize=5cm\epsfbox{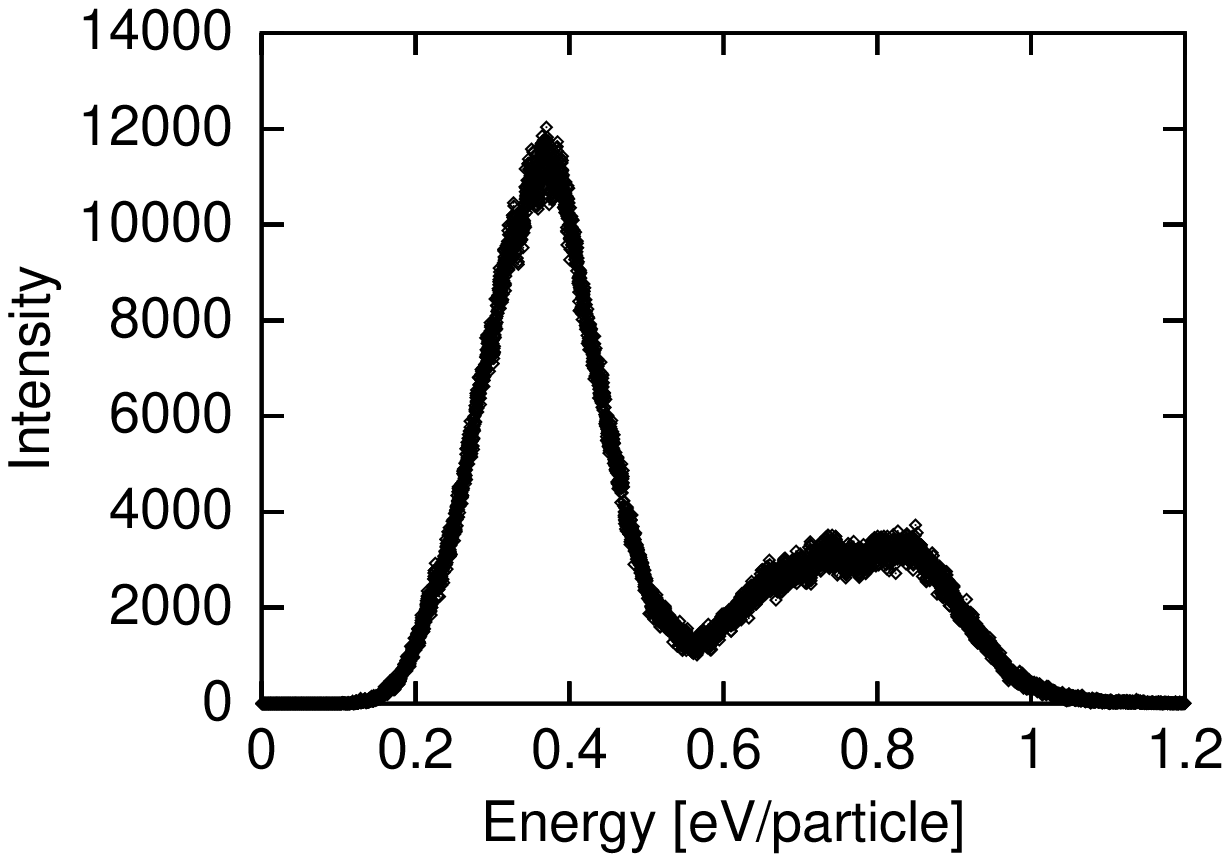}}
}
\centerline{
{\epsfxsize=6.50cm \epsfysize=5cm\epsfbox{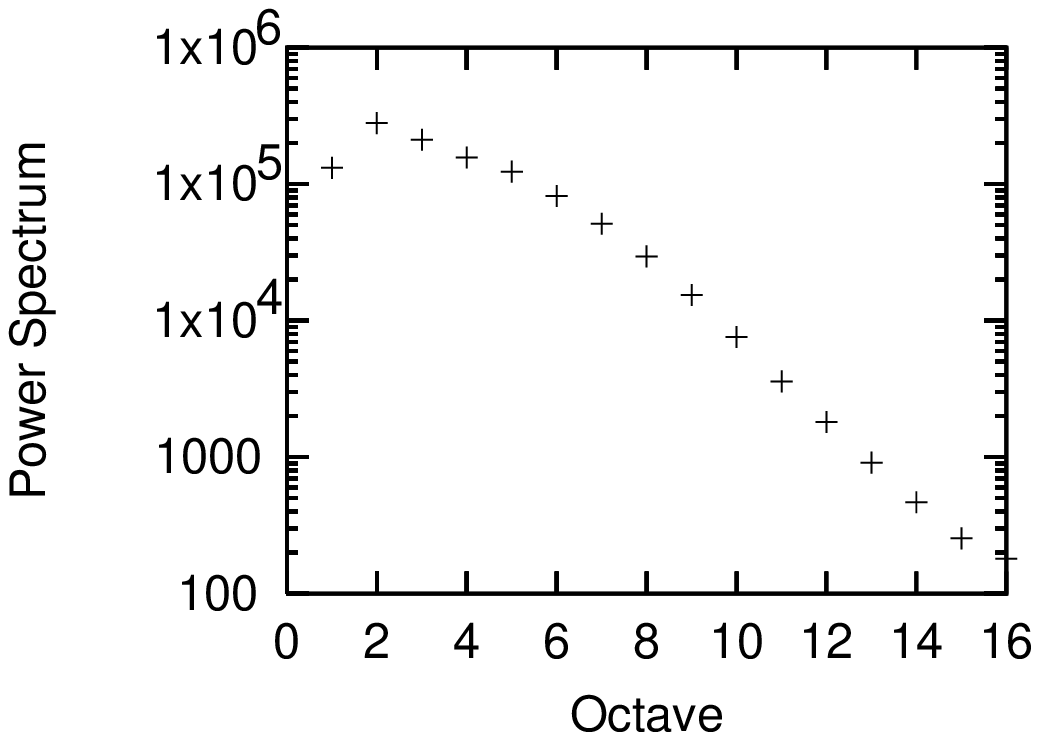}}
}
\caption{Top: a histogram of the energies at constant (high) temperature of
  order of the 
  activation energy for a transition between the two energy basins,
  depicted by the states shown in Fig.\ \ref{fig:intermittent}.  The
  left maximum corresponds to the ground state, Fig.\ 
  \ref{fig:intermittent}a, and the right to a family of low lying
  excited states, Fig.\ \ref{fig:intermittent}b). Neither peak can be
  fit by a simple gaussian, indicating a glassy nature of the ordered
  states. Bottom: The average ``power spectrum'' corresponding to the
  result shown in the top panel: the vertical axis shows the sum of
  the squares of the Fourier components of the potential energy,
  within the $n$th octave (components $2^{n-1}$ through $2^{n}-1$, for
  $n>0$), averaged over many (of order 300) spectra.  The flat low
  frequency behavior (up to about the 9th component)
  is very close to a generic\protect\cite{mike} $1/f$ noise and
  corresponds to slow fluctuations involving many particles, such as
  those yielding the transitions between the states depicted in Figs.\ 
  \ref{fig:intermittent}a and \ref{fig:intermittent}b.}
\label{fig:spectrum}
\end{figure}
Obviously, there is a band of energy states, not far (fraction of an
eV per particle) from the ground
state, which are close in energy and metastable. These states are
separated by a high barrier (the maximum of which would fall beyond
the right edge on the plot), yet are close in energy, suggesting a
\emph{rugged} energy landscape.\cite{rugged} 
Indeed, as shown earlier, formation
of a string of holes creates a barrier for adding more holes to
the string (they can be only added to the string ends). Thus,
any geometrically ordered state (say, those with denser
intra-stripe hole concentration and larger interstripe distances) 
must be separated by an energy barrier 
from other geometrically ordered states and in particular from the
ground state.

The potential energy states obtained suggest that the dynamics of
stripe motion should be strongly governed by these low lying states
and thus show a non-trivial fluctuation spectrum. Indeed, in
Fig.\ \ref{fig:spectrum}b we show the power spectrum of the
energy fluctuations for the solution described by the histogram in
Fig.\ \ref{fig:spectrum}a, and see that
the noise spectrum contains a strong $1/f$
component for approximately two and half 
decades of frequencies. This indicates 
slow fluctuations, which we ascribe to 
collective motions of melted hole string segments.

\begin{figure}
\centerline{
{\epsfxsize=7.4cm \epsfysize=5.5cm\epsfbox{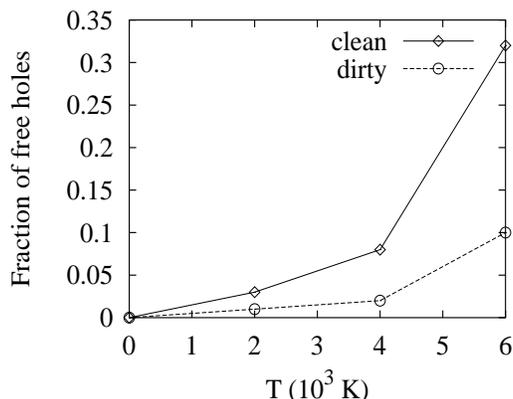}}
}
\caption{Percentage of holes which are {\em not} a part of a stripe segment,
as a function of temperature, for $B=4$eV.}
\label{fig:free_holes}
\end{figure}

Another way of characterizing the melting of stripes is by counting
``free holes:'' in Fig.\ \ref{fig:free_holes} we show the percentage
of holes which are \emph{not} in a part of an ordered pattern, as a function
of $T$. As one can see, at the transition point only a small fraction
of holes does not belong to a string segment, in agreement with our
observation that the stripes melt by rupturing into smaller segments.

Further study of the glassy dynamics of charge ordered phases in terms 
of the (free) energy landscape will be presented elsewhere.

\section{conclusions}
\label{sec:conclusions}
In summary, on employing the SDW picture of transition metal oxides,
we have studied the short-range {and}
dipolar 
{attractive forces generated by the AF fluctuations}, together
with long-range Coulomb forces. We have developed a novel numerical
technique, which enables us to treat doped hole vacancies at finite
concentration. We have studied the competition between long-range and
short-range interactions and its influence on hole
ordering in layered transition metal oxides.  We have found a rich
phase diagram for the clean system which includes a Wigner solid,
diagonal stripes, grid (loops) and a macroscopic phase separation.
For intermediate values of magnetic interaction this phase diagram is
consistent with several different experimental measurements, such as
the inelastic neutron scattering. In
addition, on adding a small, but finite amount of anisotropy to the
dipolar interaction we find that the ground state of the system of
holes is the striped phase, found in La$_{2-y-x}$Nd$_y$Sr$_x$CuO$_4$.
In the geometric phases with strong magnetic interaction strength
we have found
a large string tension for the motion of holes perpendicular to the
stripe direction. This is due to the Coulomb interaction and 
indicates strong stability of the obtained phases.

We have also found the system of holes to be quite sensitive to the
presence of charged impurities. In particular, adding out-of-plane
attractive impurities pins the holes and, for small pinning energies,
increases the melting temperature of the stripe phase, although it
does yield a finite charge correlation length. In general, charged
impurities are very effective in destroying the stripe order,
especially those residing in the same plane as the holes, regardless
of whether they are attractive or repulsive, although the stripe
phases survive as finite \emph{stripe segments} up to relatively high
impurity concentrations. This suggests that nonlinear conductivity
should be prevalent.

The resulting hole patterns are the result of frustration (competition
between short-range and long-range forces): this frustration leads to
collective motions, involving large number of particles which
ultimately lead to geometricly ordered ground states. We have also
studies the dynamics of the geometric phase formation and its melting.
We find that the dynamics is characterized by a ``glassy'' behavior in
that the energy landscape is rugged, as characterized by the
spatio-temporal intermittency of the observed behavior. More
importantly, we find that, for fixed (large) size of the magnetic
(dipolar) interaction, there are fewer number of sharper minima, while
the string tension of the stripes is larger. As a consequence, in this
case the melting of the stripe phase occurs at higher temperature with
increased doping concentration.

The energy landscape is also characterized by formation of domains,
separated by defects. This picture is in agreement with recent NMR
experiments\cite{hammel} in which small activation energies are easily
attributed to domain growth and/or motion. Thus, further study of our
model will
include the dynamics of the domain growth and their melting.

\acknowledgments

We gratefully acknowledge valuable discussions with A.\ Balatsky,
A.\ Chubukov, N.\ Curro,
J.\ Gubernatis, C.\ Hammel, G.\ Ortiz, D.\ Pines, D.\ Scalapino, 
J.\ Schmalian,
S.~White and J.~Zaanen.  A.~H.~C.~N. acknowledges support from the
Alfred P.~Sloan foundation.  Work at Los Alamos was supported by the
U.\ S.\ Department of Energy.  Work at the University of California,
Riverside, was partially supported by a Los Alamos CULAR project.
Work at Lawrence Berkeley Laboratory
was partly supported by the Director, Office of Advanced
Scientific Computing Research Division of Mathematical, Information,
and Computational Sciences of the U.S. Department of Energy under
contract number DE-AC03-76SF00098.

\appendix{}

\section{Magnetic Dipoles of the Hubbard Model}
\label{hub}

In this appendix we study the Hubbard model, Eq.\ 
(\ref{eq:hubbard}) in the SDW state.
Our approach is similar to that presented in
Refs. \onlinecite{bob,andrey}. Hence we only briefly review the
calculation leading to Eq.\ (\ref{eq:interaction}), the central
equation of the paper.

The relevant order parameter in our case is the spin density in the z direction:
\begin{eqnarray}
S^z({\bf q}) = \sum_{{\bf k},\alpha} \alpha c^{\dag}_{{\bf k}+{\bf q},\alpha}
c_{{\bf k},\alpha}\,,
\label{szq}
\end{eqnarray}
which in the SDW state has a finite 
expectation value at ${\bf q} = {\bf Q} =(\pi/a,\pi/a)$ 
because of the nesting of the half-filled Fermi surface.
In this case the mean field Hamiltonian reads
\begin{eqnarray}
H_{MF} = \sum_{{\bf k},\alpha} \epsilon_k c^{\dag}_{{\bf k},\alpha}
c_{{\bf k},\alpha} - \frac{U S N}{2} \sum_{{\bf k},\alpha}
\alpha c^{\dag}_{{\bf k}+{\bf Q},\alpha}
c_{{\bf k},\alpha}\,,
\label{mfh}
\end{eqnarray}
where
\begin{eqnarray}
S &=& \frac{1}{N} \langle S^z({\bf Q}) \rangle \,,
\nonumber
\\
\epsilon_k &=& -2 t \left(\cos(k_x a)+\cos(k_y a)\right) \, .
\label{s}
\end{eqnarray}

The Hamiltonian (\ref{mfh}) can be diagonalized immediately using 
the Bogoliubov transformation:
\begin{eqnarray}
\gamma^{c}_{{\bf k},\alpha} &=& u_{{\bf k}} c_{{\bf k},\alpha} + \alpha
v_{{\bf k}} c_{{\bf k}+{\bf Q},\alpha}
\nonumber
\\
\gamma^{v}_{{\bf k},\alpha} &=& v_{{\bf k}} c_{{\bf k},\alpha} - \alpha
u_{{\bf k}} c_{{\bf k}+{\bf Q},\alpha}
\end{eqnarray}
where
\begin{eqnarray}
u_{{\bf k}} &=& \sqrt{\frac{1}{2} \left(1+\frac{\epsilon_k}{E_k}\right)}
\nonumber
\\
v_{{\bf k}} &=& \sqrt{\frac{1}{2} \left(1-\frac{\epsilon_k}{E_k}\right)}
\nonumber
\\
E_k &=& \sqrt{\epsilon_k^2 + \Delta^2}
\nonumber
\\
\Delta &=& - \frac{U S}{2} \, .
\label{define}
\end{eqnarray}

In this case the mean-field Hamiltonian reads
\begin{eqnarray}
H_{MF} = \sum_{{\bf k},\alpha} E_k \left(\gamma^{c \dag}_{{\bf k},\alpha} 
\gamma^{c}_{{\bf k},\alpha}-\gamma^{v \dag}_{{\bf k},\alpha} 
\gamma^{v}_{{\bf k},\alpha}
\right) \,,
\label{hmfd}
\end{eqnarray}
where the sum over ${\bf k}$ is restricted to the magnetic Brillouin
zone and, at half filling, the ground state $|0\rangle$ is defined such
that
\begin{eqnarray}
\gamma^{c}_{{\bf k},\alpha} |0\rangle &=& 0
\nonumber
\\
\gamma^{v \dag}_{{\bf k},\alpha} |0\rangle &=& 0 \, .
\label{ground}
\end{eqnarray} 

Thus, at half-filling the conduction band is empty and the valence band
is separated from it by energy
$\Delta$ which is the Mott-Hubbard gap.
It is known that this theory recovers the results of the Heisenberg
model very well. Consider, for instance, the average spin density in
(\ref{szq}) in terms of the new operators (recall that the conduction
band is empty and therefore does not contribute)
\begin{eqnarray}
\langle S^z({\bf q}) \rangle &=& -2 \sum_{{\bf k},\alpha}
u_{{\bf k}+{\bf q}-{\bf Q}} v_{{\bf k}} \langle 
\gamma^{v \dag}_{{\bf k}+{\bf q}-{\bf Q},\alpha} \gamma^{v}_{{\bf k},\alpha}
\rangle \nonumber \\
&=& -4 \delta_{{\bf q},{\bf Q}} \sum_{{\bf k}}
u_{{\bf k}} v_{{\bf k}} 
= -4 \delta_{{\bf q},{\bf Q}} \sum_{{\bf k}} \frac{\Delta}{2 E_k}\,,
\label{mszq}
\end{eqnarray}
which, together with (\ref{define}), yields the gap equation:
\begin{eqnarray}
\frac{1}{N} \sum_{{\bf k}} \frac{1}{\sqrt{\epsilon_k^2+\Delta^2}}
= \frac{1}{U} \, .
\end{eqnarray} 

Since we are going to consider hole doping we can neglect the terms
involving 
the conduction band operators for temperatures $T\ll\Delta$. As shown
in 
\onlinecite{bob}
new
interactions are generated by the antiferromagnet in the presence
of the holes, given by $H_z$ and $H_{xy}$. The
non-interaction hole Hamiltonian is, from (\ref{hmfd}):
\begin{eqnarray}
H_0 = - \sum_{{\bf k},\alpha} E_k \gamma^{v \dag}_{{\bf k},\alpha} 
\gamma^{v}_{{\bf k},\alpha} \,.
\end{eqnarray}
Close to the half-filled Fermi surface one sees that the hole
mass is
\begin{eqnarray}
m_h \approx \frac{\Delta}{8 t^2 a} \, .
\end{eqnarray}
For any state $|\Psi\rangle$ of the system we can define the
hole operators as
\begin{eqnarray}
\gamma^{v}_{{\bf k},\alpha} |\Psi\rangle =
h^{\dag}_{-{\bf k},-\alpha} |\Psi\rangle
\end{eqnarray}
in which case the Hamiltonian reads
\begin{eqnarray}
H_0 = \sum_{{\bf k},\alpha} E_k h^{\dag}_{{\bf k},\alpha}
h_{{\bf k},\alpha}
\end{eqnarray}
plus unimportant constants. 

The interacting parts
of the Hamiltonian can also be written in terms of this new operators.
For instance,
\begin{eqnarray}
H_z &=& \frac{1}{N} \sum_{{\bf k},\alpha} \left[
\left(V_z({\bf Q})- V_z(2 {\bf k}-{\bf Q})/4\right) m_{k,k}^2 
- V_z(2 {\bf k}) l_{k,k}^2/4\right] h^{\dag}_{{\bf k},\alpha}
h_{{\bf k},\alpha}
\nonumber
\\
&-& \frac{1}{4 N} \sum_{{\bf k},{\bf k}'}
\left(V_z({\bf k}-{\bf k}') l_{k,k'}^2 h^{\dag}_{{\bf k},\alpha}
\sigma^z_{\alpha,\alpha'} h_{{\bf k}',\alpha'} h^{\dag}_{-{\bf k},\beta}
\sigma^z_{\beta,\beta'} h_{-{\bf k}',\beta'} \right.
\nonumber
\\
&+& \left. V_z({\bf k}-{\bf k}'+{\bf Q}) m_{k,k'}^2 h^{\dag}_{{\bf k},\alpha}
h_{{\bf k}',\alpha} h^{\dag}_{-{\bf k},\beta}
h_{-{\bf k}',\beta}\right)\,,
\label{hz}
\end{eqnarray}
where the sum over spin indices is implicit and 
\begin{eqnarray}
V_z(q) = \frac{U^2 \chi^z_0(q)}{1-U \chi^z_0(q)}
\label{vz}
\end{eqnarray}
with
\begin{eqnarray}
\chi^z_0(q,\omega) = - \frac{1}{2 N} \sum_{{\bf k}}
\left(1-\frac{\epsilon_k \epsilon_{k+q}+\Delta^2}{E_k E_{k+q}}\right)
\left(\frac{1}{\omega-E_{k+q}-E_k}-\frac{1}{\omega+E_{k+q}+E_k}\right) \, .
\label{cz}
\end{eqnarray}
A similar expression
is valid for the transverse components of the interaction
\begin{eqnarray}
H_{xy} &=& - \frac{4}{N} \sum_{{\bf k},\alpha}
(1-\alpha) \left[V_{+-}(2 {\bf k}) n^2_{k,k} -
V_{+-}({\bf Q}+2{\bf k})p^2_{k,k}\right] h^{\dag}_{{\bf k},\alpha}
h_{{\bf k},\alpha}
\nonumber
\\
&-&\frac{1}{4 N} \sum_{{\bf k},{\bf k}'}
\left[V_{+-}({\bf k}-{\bf k}') n^2_{k,k'} -
V_{+-}({\bf k}-{\bf k}'+{\bf Q})p^2_{k,k'}\right]
\nonumber
\\
&\times& 
h^{\dag}_{{\bf k},\alpha}
\sigma^+_{\alpha,\alpha'} h_{{\bf k}',\alpha'} h^{\dag}_{-{\bf k},\beta}
\sigma^-_{\beta,\beta'} h_{-{\bf k}',\beta'} \,,
\label{hxy}
\end{eqnarray}
where $V_{+-}$ is given by an expression similar to (\ref{vz})
with $\chi^z_0$ replaced by
\begin{eqnarray}
\chi^{+-}_0(q,\omega) =  = - \frac{1}{2 N} \sum_{{\bf k}}
\left(1-\frac{\epsilon_k \epsilon_{k+q}-\Delta^2}{E_k E_{k+q}}\right)
\left(\frac{1}{\omega-E_{k+q}-E_k}-\frac{1}{\omega+E_{k+q}+E_k}\right) \, .
\label{cpm}
\end{eqnarray}
Moreover, the coefficients that appear in these expressions are
defined by
\begin{eqnarray}
m_{k,k'} &=& u_k v_{k'} + v_k u_{k'}
\nonumber
\\
l_{k,k'} &=& u_k u_{k'} + v_k v_{k'}
\nonumber
\\
p_{k,k'} &=& u_k v_{k'} - v_k u_{k'}
\nonumber
\\
n_{k,k'} &=& u_k u_{k'} - v_k v_{k'} \, .
\end{eqnarray}

Observe that the interactions renormalize the dispersion of the
holes as well, that is, $E_k \to E^R_k$. 
While $H_z$ is essentially a short range attractive
interaction in the spin-symmetric channel, $H_{xy}$ has two
components: one of them is also an attractive interaction in
the spin-symmetric channel but the other is a long range dipolar
interaction which depends on the momentum of the hole. 

The Hamiltonian, which consists of terms given by Eqs.\  
(\ref{hz}) and (\ref{hxy}), is very hard to deal with. One notices,
however, that the mean field energy $E_k$ is degenerate
along the magnetic Brillouin zone. This is an artifact of
the theory and the degeneracy is broken by any small perturbation
such as the corrections discussed before or 
a next nearest hopping, $t'$, for instance. In this case, the dispersion
has a minimum at $(\pm \pi/2,\pm \pi/2)$. Thus, in order to
study the long wavelength limit of the theory it is sufficient to
focus on these points of the Brillouin zone. 
In the paper by Schrieffer, Wen and Zhang \cite{bob} 
the authors focused entirely on the $H_z$ part of the Hamiltonian
since the form factors $n_{k,k'}$ and $p_{k,k'}$ vanish at
the Brillouin zone. 
As shown by Frenkel and Hanke,\cite{david} if one keeps the leading
order in momentum we can write
\begin{eqnarray}
p_{k,k'} &\approx& \frac{t}{\Delta} |(k_x-k_x')+(k_y-k_y')|
\nonumber
\\
V_{+-}({\bf q}+{\bf Q}) &\approx& \frac{1}{t^2} \frac{1}{q^2}
\end{eqnarray}
and therefore the interaction term becomes
\begin{eqnarray}
V_{+-}({\bf q}+{\bf Q})p^2_{k,k+q} \approx 2 U \frac{(q_x+q_y)^2}{q^2}
= 2 U \left(1 + 2 \frac{q_x q_y}{q^2}\right)\,,
\end{eqnarray}
which has dipolar form.

Thus, in the first quantized
language the interactions have the form
\begin{eqnarray}
H_{I} &\approx& \left[ 
A \sigma^z({\bf r}_1) \sigma^z({\bf r}_2)
- B  \left(\sigma^+({\bf r}_1) \sigma^-({\bf r}_2)
+ \sigma^-({\bf r}_1) \sigma^+({\bf r}_2)\right)\right] \delta({\bf r}_1
-{\bf r}_2) 
\nonumber
\\
&-& C \frac{x y}{r^4}
\left(\sigma^+({\bf r}_1) \sigma^-({\bf r}_2)
+ \sigma^-({\bf r}_1) \sigma^+({\bf r}_2)\right)\,,
\label{hi}
\end{eqnarray}
where $\sigma$ are spin operators.
Notice that the singlet state $|\uparrow,\downarrow\rangle
- |\downarrow,\uparrow\rangle$, clearly minimizes the
energy of interaction between the two holes. In this case we
end up with the charge interactions only.
 
From our simulations we see
that the charge orders with some characteristic vector
${\bf K}$ such that
\begin{eqnarray}
\rho({\bf q}) = \sum_{{\bf k},\alpha} h^{\dag}_{{\bf k}+{\bf q},\alpha}
h_{{\bf k},\alpha}
\end{eqnarray}
acquires a finite expectation value at ${\bf q}= {\bf K}$.
Thus, we can always write down a mean field version of
(\ref{hi}) plus the long range Coulomb interaction 
as
\begin{eqnarray}
H_I = - \frac{\rho N}{2}  \sum_{{\bf k},\alpha} V_k 
h^{\dag}_{{\bf k} + {\bf K},\alpha} h_{{\bf k},\alpha}\,,
\label{newhi}
\end{eqnarray}
where $V_k$ has to be calculated from (\ref{hz}) and (\ref{hxy})
and
\begin{eqnarray}
\rho = \frac{1}{N} \langle \rho({\bf K}) \rangle \, .
\label{rhomf}
\end{eqnarray}
Observe that in this case the Brillouin zone is further reduced
and we can define new operators
\begin{eqnarray}
d^+_{{\bf k},\alpha} = w_k h_{{\bf k},\alpha} + t_k 
h^{\dag}_{{\bf k} + {\bf K},\alpha}
\nonumber
\\
d^-_{{\bf k},\alpha} = t_k h_{{\bf k},\alpha} - w_k
h^{\dag}_{{\bf k} + {\bf K},\alpha}
\end{eqnarray}
with
\begin{eqnarray}
w_k &=& \sqrt{\frac{1}{2} \left(1+\frac{E^R_k}{E^T_k}\right)}
\nonumber
\\
t_k &=& \sqrt{\frac{1}{2} \left(1-\frac{E^R_k}{E^T_k}\right)}
\nonumber
\\
E^T_k &=& \sqrt{(E^R_k)^2 + \frac{(\rho V_k)^2}{4}}
\end{eqnarray}
and the Hamiltonian is diagonal
\begin{eqnarray}
H = \sum_{{\bf k},\alpha} E^T_k \left(d^{+ \dag}_{{\bf k},\alpha}
d^+_{{\bf k},\alpha} - d^{- \dag}_{{\bf k},\alpha} d^{-}_{{\bf k},\alpha}
\right) \,,
\end{eqnarray}
where the sum is done in the new Brillouin zone.
Self-consistency requires that
\begin{eqnarray}
\rho = \frac{1}{N} \sum_{{\bf k},\alpha} w_k t_k 
\left(\langle d^{+ \dag}_{{\bf k},\alpha}
d^+_{{\bf k},\alpha} \rangle - \langle 
d^{- \dag}_{{\bf k},\alpha} d^{-}_{{\bf k},\alpha} \rangle
\right)
\end{eqnarray}
which can be evaluated for any hole-filling.

Now let us go back to the issue of magnetization which is important
for neutron scattering. From (\ref{mszq}) one has
\begin{eqnarray}
\langle S^z({\bf q}) \rangle &=& -2 \sum_{{\bf k},\alpha}
u_{{\bf k}+{\bf q}-{\bf Q}} v_{{\bf k}} \langle 
\gamma^{v \dag}_{{\bf k}+{\bf q}-{\bf Q},\alpha} \gamma^{v}_{{\bf k},\alpha}
\rangle
\nonumber
\\
&=& -2 \sum_{{\bf k},\alpha}
u_{{\bf k}+{\bf q}-{\bf Q}} v_{{\bf k}} \langle 
h_{{\bf k}+{\bf q}-{\bf Q},\alpha} h^{\dag}_{{\bf k},\alpha}
\rangle
\nonumber
\\
&=& -2 \delta_{{\bf q},{\bf Q}+{\bf K}} \sum_{{\bf k},\alpha}
u_{{\bf k}+ {\bf K}}  v_{{\bf k}} \langle h_{{\bf k} + {\bf K},\alpha}
h^{\dag}_{{\bf k},\alpha}
\rangle
\nonumber
\\
&-&2 \delta_{{\bf q},{\bf Q}+{\bf K}} \sum_{{\bf k},\alpha}
u_{{\bf k}+ {\bf K}}  v_{{\bf k}} t_k w_k
\left(\langle d^{+}_{{\bf k},\alpha}
d^{+ \dag}_{{\bf k},\alpha} \rangle - \langle 
d^{-}_{{\bf k},\alpha} d^{- \dag}_{{\bf k},\alpha} \rangle
\right)
\end{eqnarray}
and one sees that the magnetization is now peaked around
${\bf Q}+{\bf K}$ instead of ${\bf Q}$. For stripes 
aligned along the ${\bf y}$ direction this
is possible of course when
\begin{eqnarray}
{\bf K} = \pm \frac{2 \pi}{\ell} {\bf x}\,,
\end{eqnarray}
where $\ell$ is the inter-stripe distance. 

\section{Dipoles of the $t-J$ model}
\label{tj}

Let us consider the SDW theory of
the $t-J$ model {\it \`a la} Shraiman and Siggia,\cite{ss} described by
\begin{eqnarray}
H=-t \sum_{\langle i,j \rangle} c^{\dag}_{i,\sigma} c_{j,\sigma} + h.c.
+ J \sum_{\langle i,j \rangle} {\bf S}_i \cdot {\bf S}_j
\label{tjmod}
\end{eqnarray}
and use the slave fermion representation
\begin{eqnarray}
c_{i,\sigma} = \psi^{\dag}_{\alpha,i} z_{\alpha,i,\sigma}\,,
\label{sf}
\end{eqnarray}
where $\psi^{\dag}_{\alpha,i}$ creates a hole (fermion) on
a site $i$ in the sublattice $\alpha=A,B$ (which labels the
"spin" of the hole) and $z_{\alpha,i,\sigma}$
is a Schwinger boson on the same sublattice (spin wave). 
In order to obtain the dynamics of the holes alone we trace
out the spin-wave degrees of freedom. The static part of
the interaction is\cite{ss}
\begin{eqnarray}
H_{SS} = -\frac{1}{N} \sum_{{\bf k},{\bf k'},{\bf q}} 
V({\bf k},{\bf k'},{\bf q}) \psi^{\dag}_{A}({\bf k})
\psi_{B}({\bf k}+{\bf q}) \psi^{\dag}_{B}({\bf k'}+{\bf q})
\psi_{A}({\bf k'})\,,
\label{ssh}
\end{eqnarray}
where the momentum sum is restricted to the magnetic Brillouin zone and
\begin{eqnarray}
V({\bf k},{\bf k'},{\bf q}) &=& - g \left[\frac{\left(\lambda_{{\bf k}}
- \lambda_{{\bf k}+{\bf q}}\right)\left(\lambda_{{\bf k'}}
- \lambda_{{\bf k'}+{\bf q}}\right)}{1-\lambda_{{\bf q}}}\right.
\nonumber
\\
&+& \left. \frac{\left(\lambda_{{\bf k}}
+ \lambda_{{\bf k}+{\bf q}}\right)\left(\lambda_{{\bf k'}}
+ \lambda_{{\bf k'}+{\bf q}}\right)}{1+\lambda_{{\bf q}}}\right]
\label{vkkq}
\end{eqnarray}
with
\begin{eqnarray}
\lambda_{{\bf q}} = \frac{1}{2} \left(\cos(q_x) + \cos(q_y)\right) \, .
\label{lambda}
\end{eqnarray}
The coupling constant $g$ is a function of $t$ and $J$. In the strong
coupling limit ($t<<J$) $g \approx 8 t^2/J$, while in the weak
coupling limit ($t>>J$) we have $g \approx J$ \cite{david}.

\begin{figure}
\epsfxsize=3.2 truein
\centerline{\epsffile{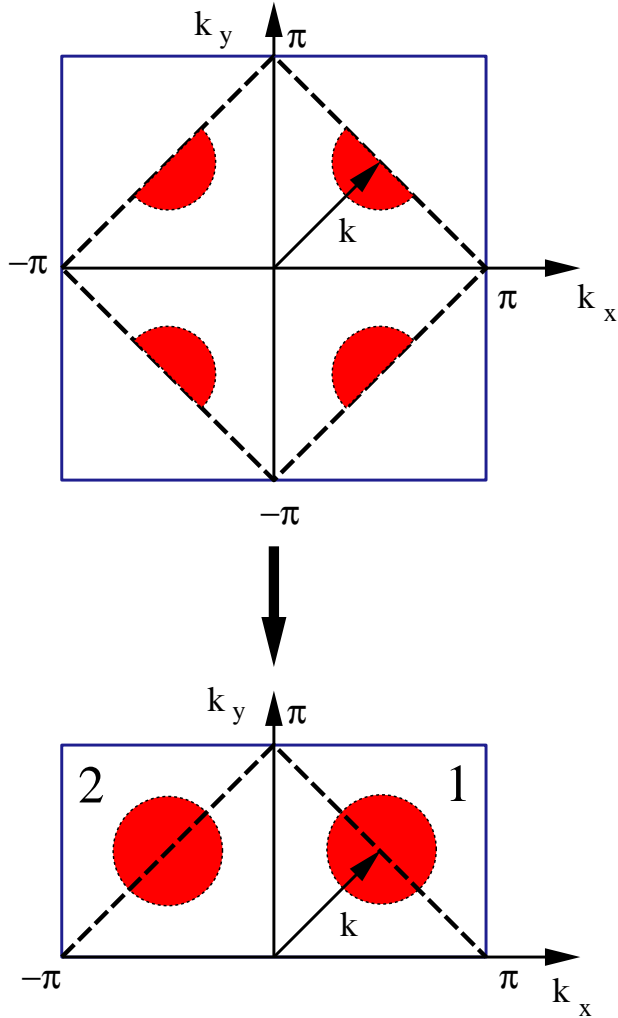}}
\caption{Choice of the BZ.}
\label{fold}
\end{figure}

Observe that (\ref{ssh}) does not have a kinetic term for the holes.
The kinetic energy has to be obtained from the hole self-energy
at zero frequency and can be written as
\begin{eqnarray}
H_0 = \sum_{{\bf k},\alpha=A,B} \epsilon_{{\bf k}} 
\psi^{\dag}_{\alpha}({\bf k}) \psi_{\alpha}({\bf k})\,,
\label{ek}
\end{eqnarray}
where $\epsilon_{{\bf k}}$ has a minimum at $(\pm \pi/2,\pm \pi/2)$
\cite{ss}. For a low density of holes these are the only points
of interest and therefore we can look at the interaction
(\ref{vkkq}) strength close to these points. Observe that
at these points we have $\lambda_{{\bf q}} \to 0$ and therefore
the interactions are dominated by the first term in (\ref{vkkq})
which describes the fluctuations of the staggered magnetization
(with characteristic wave-vector ${\bf Q} = (\pi,\pi)$). The 
second term describes the fluctuations of the homogeneous magnetization
(${\bf q}=0$) which is not of direct interest here.
In this case, for ${\bf k}$ and ${\bf k'}$
close $(\pm \pi/2,\pm \pi/2)$ to the interaction can be approximated by
\begin{eqnarray}
V({\bf k},{\bf k'},{\bf q}) \approx -g \frac{\lambda_{{\bf k}+{\bf q}}
\lambda_{{\bf k'}+{\bf q}}}{1-\lambda_{{\bf q}}} \, .
\label{vkkqap}
\end{eqnarray} 
The problem can be further simplified if one works with the upper
half-part of the original BZ instead of the magnetic BZ, as shown
in Fig.\ref{fold}. This can be accomplished by a shift of the
lower part of the BZ by ${\bf Q}$.

Moreover, working in the long wave-length limit, that is,
with ${\bf q} \to 0$, one sees
that there are four values of $V({\bf k},{\bf k'},{\bf q})$
of relevance: 1) when ${\bf k} = {\bf k'} = {\bf Q}/2$ 
\begin{eqnarray}
V_{11}({\bf q}) \approx -g \frac{(q_x+q_y)^2}{q^2} \, ;
\label{v11}
\end{eqnarray}
2) when ${\bf k} = {\bf k'} = {\bf Q}^*/2$ (where
${\bf Q}^* = (-\pi,\pi)$) and
\begin{eqnarray}
V_{22}({\bf q}) \approx -g \frac{(q_x-q_y)^2}{q^2} \, ;
\label{v22}
\end{eqnarray}
3)when ${\bf k} = {\bf Q}/2$ and ${\bf k'} = {\bf Q}^*/2$
\begin{eqnarray}
V_{12}({\bf q}) \approx g \frac{q^2_x-q^2_y}{q^2} \, ;
\label{v12}
\end{eqnarray}
4)and finally ${\bf k} = {\bf Q}^*/2$ and ${\bf k'} = {\bf Q}/2$
\begin{eqnarray}
V_{21}({\bf q}) = V_{12}({\bf q})  \, .
\label{v21}
\end{eqnarray}
We can now split the sums in (\ref{ssh}) to the regions around
these two points and introduce a cut-off in the momentum sum
$\Lambda$ such that $q<<\Lambda<<\pi$. In this case the Hilbert
space of the problem is divided into two different sub-Hilbert
spaces and the hole operator can be rewritten as
\begin{eqnarray}
\psi_{\alpha,i} &=& \sum_{{\bf k}} e^{i {\bf k} \cdot {\bf r}_i} 
\psi_{\alpha}({\bf k})
\nonumber
\\
&\approx& \psi_{\alpha,i,1} \cos\left(\frac{{\bf Q}}{2} \cdot {\bf r}\right)
+ \psi_{\alpha,i,2} \cos\left(\frac{{\bf Q}^*}{2} \cdot {\bf r}\right)\,,
\end{eqnarray}
where
\begin{eqnarray}
\psi_{\alpha,i,1} &\approx& \sum_{{\bf q}} e^{i {\bf q} \cdot {\bf r}_i}
\psi_{\alpha}(\frac{{\bf Q}}{2}+{\bf q})
\nonumber
\\
\psi_{\alpha,i,2} &\approx& \sum_{{\bf q}} e^{i {\bf q} \cdot {\bf r}_i}
\psi_{\alpha}(\frac{{\bf Q}^*}{2}+{\bf q}) \, .
\end{eqnarray}

It is convenient to define the operator
\begin{eqnarray}
p_a({\bf q}) &=& \sum_{{\bf k}} \psi^{\dag}_{B,a}({\bf k}+{\bf q})
\psi_{A,a}({\bf k})
\nonumber
\\
p^{\dag}_a({\bf q}) &=& \sum_{{\bf k}} \psi^{\dag}_{A,a}({\bf k})
\psi_{B,a}({\bf k}+{\bf q}) = \sum_{{\bf k}} \psi^{\dag}_{A,a}({\bf k}-{\bf q})
\psi_{B,a}({\bf k})
\label{pa}
\end{eqnarray}
with $a=1,2$. Using these new operators and results (\ref{v11})-(\ref{v21}),
we can rewrite the Hamiltonian (\ref{ssh}) as
\begin{eqnarray}
H_{SS} &\approx& -\frac{g}{N} \sum_{{\bf q}}
\frac{1}{q^2} \left[q^2 \left(p^{\dag}_1({\bf q}) p_1({\bf q})
+ p^{\dag}_2({\bf q}) p_2({\bf q})\right) \right. 
\nonumber
\\
&-& \left. (q_x^2-q_y^2) \left(p^{\dag}_1({\bf q}) p_2({\bf q})
+ p^{\dag}_2({\bf q}) p_1({\bf q})\right)
+ 2 q_x q_y \left(p^{\dag}_1({\bf q}) p_1({\bf q})
- p^{\dag}_2({\bf q}) p_2({\bf q})\right)\right]\,.
\label{hss1}
\end{eqnarray}
This does not have a very transparent form. In order
to see that this Hamiltonian has the form of a dipole-dipole interaction
we define the vector operator
\begin{eqnarray}
{\bf D}({\bf q}) &=& \frac{1}{\sqrt{2}} 
\left(p_1({\bf q})-p_2({\bf q}),p_1({\bf q})+p_2({\bf q}) \right)
\nonumber
\\
{\bf D}^{\dag}({\bf q}) &=& \frac{1}{\sqrt{2}}
\left(p^{\dag}_1(-{\bf q})-p^{\dag}_2(-{\bf q}),p^{\dag}_1(-{\bf q})+
p^{\dag}_2(-{\bf q}) \right) 
\label{bigd}
\end{eqnarray}
and rewrite (\ref{hss1}) as
\begin{eqnarray}
H_{SS} &\approx& -\frac{g}{N} \sum_{{\bf q}}
\frac{1}{q^2} \left[q^2 {\bf D}^{\dag}({\bf q}) \cdot {\bf D}({\bf q})
\right.
\nonumber
\\
&+& \left. {\bf D}^{\dag}(-{\bf q}) \cdot {\bf D}({\bf q}) 
- 2 \frac{\left({\bf D}^{\dag}(-{\bf q}) \cdot {\bf q}\right)
\left({\bf D}({\bf q}) \cdot {\bf q}\right)}{q^2}\right] \, .
\label{almost}
\end{eqnarray}
Observe that the first term in (\ref{almost}) is q-independent and
will lead to a local interaction which has the usual scalar form.
We are interested in the second part of the Hamiltonian. Defining
the Fourier transform
\begin{eqnarray}
D({\bf r}) = \sum_{{\bf q}} e^{i {\bf q} \cdot {\bf r}} D({\bf q})
\end{eqnarray}
the second term in (\ref{almost}) acquires the required form
\begin{eqnarray}
H_{dd} = g \int d{\bf r} \int d{\bf r'} \frac{1}{({\bf r}-{\bf r'})^2} 
\left\{{\bf D}^{\dag}({\bf r}) \cdot {\bf D}({\bf r'})
-2 \frac{\left({\bf D}^{\dag}({\bf r}) \cdot {\bf r}\right)
\left({\bf D}({\bf r'}) \cdot {\bf r'}\right)}{({\bf r}-{\bf
  r'})^2}\right\}\,, 
\label{hdd}
\end{eqnarray}
which is the second quantized form of the magnetic dipole-dipole
interaction. 

In order to put this Hamiltonian in 
a form involving the magnetic dipole defined by Shraimann and Siggia
consider their definition \cite{ss}:
\begin{eqnarray}
P_{\mu,\alpha}({\bf q}) = \sum_{{\bf k},a,b} 
\sin(k_{\alpha}) \psi^{\dag}_{a}({\bf k}+{\bf q})
\tau^{\mu}_{a,b} \psi_{b}({\bf k})\,,
\label{pma}
\end{eqnarray}
where $\mu=x,y,z$ refers to indices of the Pauli matrices
$\tau^{\mu}_{a,b}$ and therefore act in the sub-space of the
sublattices $A$ and $B$ and $\alpha=x,y$ refers to the space
indices. In particular, the lowering and raising operators
associated with this magnetic dipole operators have the
interesting form
\begin{eqnarray}
P_{-,\alpha}({\bf q}) &=& 2 \sum_{{\bf k}} \sin(k_{\alpha}) \psi^{\dag}_{B}(
{\bf k}+{\bf q}) \psi_{A}({\bf k})
\nonumber
\\
P_{+,\alpha}({\bf q}) &=& 2 \sum_{{\bf k}} \sin(k_{\alpha}) \psi^{\dag}_{A}(
{\bf k}+{\bf q}) \psi_{B}({\bf k}) \,.
\label{pmm}
\end{eqnarray}
In the approximation we are employing we can split the
summation in (\ref{pmm}) around ${\bf Q}$ and ${\bf Q}^*$ in order
to get (this can be done because the sine function is smooth around
these two points)
\begin{eqnarray}
P_{-,x}({\bf q}) &\approx& 2 \sum_{{\bf k}} \left[\psi^{\dag}_{B,1}(
{\bf k}+{\bf q}) \psi_{A,1}({\bf k})-\psi^{\dag}_{B,2}(
{\bf k}+{\bf q}) \psi_{A,2}({\bf k})\right]
\nonumber
\\
P_{-,y}({\bf q}) &\approx& 2 \sum_{{\bf k}} \left[\psi^{\dag}_{B,1}(
{\bf k}+{\bf q}) \psi_{A,1}({\bf k})+\psi^{\dag}_{B,2}(
{\bf k}+{\bf q}) \psi_{A,2}({\bf k})\right]
\nonumber
\\
P_{+,x}({\bf q}) &\approx& 2 \sum_{{\bf k}} \left[\psi^{\dag}_{A,1}(
{\bf k}+{\bf q}) \psi_{B,1}({\bf k})-\psi^{\dag}_{A,2}(
{\bf k}+{\bf q}) \psi_{B,2}({\bf k})\right]
\nonumber
\\
P_{+,x}({\bf q}) &\approx& 2 \sum_{{\bf k}} \left[\psi^{\dag}_{A,1}(
{\bf k}+{\bf q}) \psi_{B,1}({\bf k})+\psi^{\dag}_{A,2}(
{\bf k}+{\bf q}) \psi_{B,2}({\bf k})\right] \,,
\label{pmmap}
\end{eqnarray}
where the signs come from value of the sines around the
two points in the FS. 
These dipole operators can be also
written trivially  in terms of the operators in (\ref{pa}):
\begin{eqnarray}
P_{-,x}({\bf q}) &=& 2 \left(p_{1}({\bf q}) - p_{2}({\bf q})\right)
\nonumber
\\
P_{-,y}({\bf q}) &=& 2 \left(p_{1}({\bf q}) + p_{2}({\bf q})\right)
\end{eqnarray}
and so on. By direct comparison with (\ref{bigd}) one
finds
\begin{eqnarray}
{\bf D}({\bf q}) = \frac{1}{2 \sqrt{2}} 
\left(P_{-,x}({\bf q}),P_{-,y}({\bf q})\right) \,,
\end{eqnarray}
which makes clear the connection. On taking the 
Fourier transform of the magnetic dipole operators back
to real space one finds, for instance,
\begin{eqnarray}
D_{x,i} &=& \frac{1}{\sqrt{2}}\left(
\psi^{\dag}_{B,1,i} \psi_{A,1,i}-\psi^{\dag}_{B,2,i} \psi_{A,2,i}\right)
\nonumber
\\
D_{y,i} &=& \frac{1}{\sqrt{2}}\left(
\psi^{\dag}_{B,1,i} \psi_{A,1,i}+\psi^{\dag}_{B,2,i} \psi_{A,2,i}\right)\,.
\end{eqnarray}
This explicitely justifies our earlier claim that the dipolar
interaction is due to the coherent hoping of holes between two
different sublattices (at the same position in space).


\begin{references}
\bibitem{SK} We are thankful to S.\ Kivelson for attracting our
attention to this problem.
\bibitem{ferro-vertex} The simulations show the presence of
  topological defects in the phase of the magnetic dipole moments, i.e.,
  the total phase in a loop may be shifted by $2\pi$.
\end{references}
\end{document}